\documentclass[11pt]{article}
\usepackage[pdftex]{graphicx}
   \graphicspath{{figuresPDF/}}
   \DeclareGraphicsExtensions{.pdf,.jpeg,.png, .eps}

\usepackage{url}
\usepackage{hyperref}
\usepackage{amsfonts,amssymb,dsfont}
\usepackage{amsmath, bm}
\usepackage{algorithm, algorithmic}
\usepackage{color}
\usepackage{authblk, comment}

\usepackage[authoryear, round]{natbib}

\def\vk{\vskip0.1cm}

\newcommand{\bw}{\mathbf{w}}

\newcommand{\bsv}{\boldsymbol{v}}

\newcommand{\bsw}{\mathbf{w}}

\newcommand{\bsx}{\boldsymbol{x}}
\newcommand{\bsX}{\boldsymbol{X}}

\newcommand{\cX}{\mathcal{X}}
\newcommand{\cY}{\mathcal{Y}}

\newcommand{\cD}{\mathcal{D}}

\newcommand{\bsbeta}{\boldsymbol{\beta}}

\newcommand{\bstheta}{\boldsymbol{\theta}}

\newcommand{\E}{\mathbb{E}}

\newcommand{\Pro}{\mathbb{P}}
\newcommand{\R}{\mathbb{R}}

\newcommand{\N}{\mathbb{N}}

\begin{document}
    
\title{Regularized Estimation and Feature Selection in \\Mixtures of Generalized Linear Experts}      

\author[1,2]{Thin Nguyen-Van\thanks{Email: \texttt{nvthin@hcmus.edu.vn}}}
\author[3]{Faicel Chamroukhi\thanks{Email: \texttt{faicel.chamroukhi@irt-systemx.fr}}}
\author[1]{Ha Hoang Van\thanks{Email: \texttt{hvha@hcmus.edu.vn}}}
\author[4]{Bao Tuyen Huynh\thanks{Email: \texttt{tuyenhb@dlu.edu.vn}}}

\affil[1]{Faculty of Mathematics and Computer Science, University of Science, Ho Chi Minh City, Vietnam}
\affil[2]{Vietnam National University, Ho Chi Minh City, Vietnam}
\affil[3]{IRT SystemX, 91120 Palaiseau, France}
\affil[4]{Faculty of Mathematics and Computer Science, Dalat University, Vietnam}

\maketitle

\begin{abstract}
Mixtures of experts (MoE) are conditional mixture models in which both the mixing proportions and the component densities depend on the predictors, and are widely used for regression, classification and model-based clustering of heterogeneous data. Fitting MoE by maximum likelihood becomes unstable, and sometimes infeasible, when the predictors are numerous or correlated. We propose a regularized maximum likelihood framework for simultaneous parameter estimation and feature selection in MoE whose experts belong to the generalized linear model family, covering Gaussian, Poisson and multinomial responses within a single formulation. Sparsity is induced in both the gating network and the experts through $\ell_1$ penalties, and the penalized log-likelihood is maximized by a proximal Newton-EM algorithm whose M-step reduces to weighted Lasso problems with closed-form coordinate-ascent updates. Unlike existing penalized MoE procedures, the algorithm requires neither a local quadratic approximation of the penalty nor any matrix inversion, it returns exactly sparse estimates without thresholding, and a proximal Newton-type variant guarantees a monotone increase of the penalized objective at every iteration. On simulated data and five real data sets, the method recovers the actual sparsity support and delivers prediction and clustering accuracy that is competitive with, and often better than, state-of-the-art regularized MoE.
The source codes of our developed algorithms and their documentation are publicly available on Github\footnote{\url{https://github.com/nv-thin/GLM-RMoE}}.
\end{abstract}


\renewcommand{\baselinestretch}{1.5}
\normalsize

\section{Introduction and related work}

Mixtures-of-experts (MoE) models introduced by \cite{Jac91}, including hierarchical MoE \cite{Jor94}, have demonstrated strong performance for statistical modeling of heterogeneous data across various statistical learning tasks such as regression, clustering, and classification.  
MoE belong to the family of mixture models \cite{McLachlan2000FMM} and are fully conditional mixtures in which both the mixing proportions (gating network) and the component densities (experts network) depend on the input variables.  
This conditional formulation enables MoE to represent complex data distributions more effectively than standard unconditional mixture models.  
The statistical inference and numerical aspects of (hierarchical) MoE have been investigated in \cite{Jor94,Jia99,Jia99b,Jiang2000}, and recent extensions for modeling and clustering heterogeneous regression data, possibly contaminated with asymmetric noise or outliers, include \citep{Chamroukhi-TMoE-16, Chamroukhi-SNMoE, Nguyen-LMoE-14, Chamroukhi-STMoE-17}.  
More recently, MoE models have been extended to functional data analysis and Bayesian modeling, considerably broadening their applicability to complex heterogeneous data structures \citep{Chamroukhi2024FME,Zhang2023BayesianMoE}.
A general overview of MoE models and their applications can be found in \cite{Yuk12,NguyenChamroukhi-MoE}.\vk  

While maximum-likelihood estimation (MLE) is widely used to fit MoE, studying MoE in high-dimensional settings remains challenging due to the well-known limitations of the ML estimator.  
Recent work such as \cite{Nguyen2022NonAsymptotic} proposed a non-asymptotic penalization framework for model selection in high-dimensional MoE, further highlighting the difficulty of performing inference when both the number of features and the model complexity grow.  
In such settings, features are often correlated and only a smaller subset is truly informative, which leads to numerical instability and potentially unbounded parameter estimates.  
This behavior is well documented in regression, particularly in multinomial regression; see \cite{Park2007b,Bunea2008}.  
Consequently, estimating MoE parameters with moderate to large numbers of features and mixture components can be difficult.  
To avoid singularities and degeneracies of the MLE \cite{Stephens1997, Fraley2007}, regularization provides a principled solution by imposing sparsity and stabilizing estimation.\vk  

Regularization for mixture regression models has been studied in several directions.  
In mixture of linear regressions (MLR), \cite{Kha07} introduced regularized MLE procedures (MIXLASSO, MIXHARD, MIXSCAD) and established asymptotic properties for various penalties; this was later extended to the MoE setting in \cite{Kha10}, who proved root-$n$ consistency and oracle properties for Lasso and SCAD within a penalized MoE framework.  
However, their approach relies on an approximation of the penalty function and requires Newton--Raphson updates involving matrix inversion, which becomes computationally expensive in high-dimensional scenarios.
More recently, \cite{Khalili2025SparseMoE} developed a sparse MoE estimation framework based on a modified EM algorithm and proximal-gradient optimization while establishing consistency properties in high-dimensional settings.
To alleviate these computational bottlenecks, \cite{Cham18} proposed a regularized MLE approach for MoE that does not require penalty approximations, developing hybrid EM/MM and coordinate ascent algorithms that avoid matrix inversions via univariate parameter updates.
\cite{Per14} considered MoE with multinomial experts and proposed an EM algorithm based on inverting the soft-max function, but provided no guarantee that their EM iterations increase the penalized objective function.  
To address challenges in updating gating-network parameters, \cite{Jiang2018} developed a penalized likelihood method for localized MoE \citep{Xu1995}, although the normal update of local covariance matrices in the M-step limits its applicability to large-scale problems. Related developments on penalized EM algorithms for high-dimensional mixture regression have also established strong statistical guarantees and demonstrated the effectiveness of combining regularization with efficient optimization techniques \citep{Wang2024PenalizedEM}.\vk  

In this paper, we propose an efficient regularized estimation and feature-selection framework for MoE models, considering three common generalized linear expert families.  
We introduce a proximal Newton--EM algorithm for maximizing the $\ell_1$-penalized log-likelihood, where the M-step is solved using a proximal Newton-type method.  
Our approach is motivated by the success of proximal Newton methods for sparse composite optimization \citep{Lee14}, which effectively combine second-order information with sparsity-inducing regularization and exhibit fast local convergence.
This reduces parameter updates to solving weighted quadratic Lasso problems, which can be efficiently handled using coordinate ascent techniques.  
Unlike methods relying on approximated penalties or thresholding, our approach directly maximizes the penalized objective and automatically yields sparse solutions.  
We demonstrate that the proposed method performs well in both moderate- and high-dimensional scenarios and outperforms competitive state-of-the-art regularized MoE models on a variety of simulated and real data sets.\vk  

The remainder of this paper is organized as follows.  
Section \ref{sec: MoE} introduces the MoE formulation for heterogeneous data and the MLE procedure.  
Section \ref{sec: RMoE} presents the proposed penalized likelihood framework and the EM-based proximal Newton algorithm.  
Section \ref{sec: Experiments} reports extensive experiments on simulated and real data.  
Finally, Section \ref{Sec:Con} concludes the paper and discusses potential future directions.
\section{Mixture-of-Experts and Maximum Likelihood Estimation}
\label{sec: MoE}
Let $((\bsX_1,Y_1),\ldots,(\bsX_n,Y_n))$ be a random sample of $n$  independent and identically distributed (i.i.d) pairs $(\bsX_i,Y_i)$, ($i=1,\ldots, n$) 
where $Y_i\in \cY \subset \R$ is the $i$th response  given some vector of $p\in\mathbb{N}$ predictors $\bsX_i \in \cX \subset \R^p$,  with $\cY$ denoting the response space appropriate to the distribution under consideration (Gaussian, Poisson, or multinomial).
We consider the MoE modeling  for the analysis of a heterogeneous set of such data.
Let $\cD = ((\bsx_1,y_1),\ldots,(\bsx_n,y_n))$ be an observed data sample. 

\subsection{The MoE model}
The mixture-of-experts model assumes that the observed pairs $(\bsx,y)$ are generated from $K\in \N$ (possibly unknown) parametric probability density components (the experts) $p_z(y|\bsx;\bstheta)$, $z\in [K]=\{1,\ldots,K\}$, governed by a gating network $\pi_z(\bsx;\bw)$ represented by a hidden multinomial random variable $Z\in [K]$ that indicates the expert to which a particular observed pair belongs. 
The generative process of the data hence assumes the following hierarchical representation. 
Given the predictor or the input $\bsx_i$, the multinomial variable $Z_i$ is generated according to the multinomial distribution:
\begin{equation}
Z_i|\bsx_i \sim \text{Mult}(1;\pi_1(\bsx_i;\bw),\hdots,\pi_K(\bsx_i;\bw)) 
\label{eq.Z generation in MoE}
\end{equation} where each of the probabilities $\pi_{z_i}(\bsx_i;\bw) = \mathbb{P}(Z_i = z_i|\bsX_i=\bsx_i)$ is given by the gating network. 
Then, conditional on the hidden variable $Z_i = z_i$ and $\bsx_i$, the observed random variable $Y_i$ is assumed to be generated from the expert $z_i$ its distribution is $p_{z_i}(y_i|\bsx_i;\bstheta_{z_i})$, that is:
\begin{equation*}
Y_i|Z_i = z_i, \bsX_i=\bsx_i   \sim p_{z_i}(y_i|\bsx_i;\bstheta_{z_i})
\end{equation*}
where $p_{z_i}(y_i|\bsx_i;\bstheta_{z_i}) = p(y_i|Z_i = z_i,\bsX_i=\bsx_i;\bstheta_{z_i})$ is the probability density or the probability mass function of the expert $z_i$ depending on the nature of the data ($\bsx,y$) within the group $z_i$. 
The gating network which gives the probabilities in (\ref{eq.Z generation in MoE}) is defined by the distribution of the hidden variable $Z$ given the predictor $\bsx$, i.e., $\pi_k(\bsx;\bw) = \Pro(Z=k|\bsX=\bsx;\bw)$, is in general given by gating softmax functions of the form:
\begin{eqnarray*}
\pi_k(\bsx_i;\bw) = \Pro(Z_i=k|\bsX_i=\bsx_i;\bw)
 = \frac{\exp(w_{k0}+\bsx_i^\top \bsw_k)}{1 + \sum\limits_{l=1}^{K-1}\exp(w_{l0}+\bsx_i^\top \bsw_l)}
\end{eqnarray*}for $k=1,\hdots,K-1$ with $\bw = (\bw^\top_1, \ldots, \bw_{K-1}^\top)^\top$ and $\bw_k = (w_{k0}, \bsw^\top_k)^\top \in \R^{p+1}$ such that  $\bw_K = \mathbf{0}$ is set to the null vector for identifiability \citep{Jia99}.
Hence, formally, the MoE is defined by the following semi-parametric probability density (or mass) function: %
\begin{equation}
p(y_i|\bsx_i; \bstheta) = \sum_{k=1}^K \pi_k(\bsx_i;\bw) p_k(y_i|\bsx_i;\bstheta_k)
\label{eq.MoE}
\end{equation}that is parameterized by the parameter vector defined by 
$\bstheta = (\bw^\top_1,\ldots,\bw^\top_{K-1},\bstheta^\top_1,\ldots,\bstheta^\top_K)^\top \in \R^{\nu_{\bstheta}}$ ($\nu_{\bstheta} \in \N$) 
where $\bstheta_k$ ($k=1,\ldots,K$) is the parameter vector of the $k$th expert.\vk
For a complete account of MoE, types of gating networks and expert networks, the reader can be referred to \cite{NguyenChamroukhi-MoE}. 

\subsection{Maximum likelihood parameter estimation}
\label{ssec: EM-MoE}
Given an observed data sample $\cD = ((\bsx_1,y_1),\ldots,(\bsx_n, y_n))$ generated from the MoE model (\ref{eq.MoE}), the unknown parameter vector $\bstheta$ is commonly estimated by maximizing the observed data log-likelihood
\begin{equation}
L(\bstheta) = \sum_{i=1}^{n}\log\sum_{k=1}^{K} \pi_k(\bsx_i;\bw) p_k(y_{i}|\bsx_i;\bstheta_{k})
\label{eq.log-lik MoE}
\end{equation}
by using the EM algorithm \citep{Dem77, Jac91} which allows to iteratively find an appropriate local maximizer of the log-likelihood function (\ref{eq.log-lik MoE}). 
\cite{Jiang2000} studied statistical estimation and numerical computations in (hierarchical) MoE models.\vk
However, it is well-known that the MLE can be unstable or even infeasible in high-dimension due to possibly redundant and correlated features. In some cases, such as multinomial model, this task becomes a challenge since the log-likelihood function becomes singular. In such a context, a regularization of the MLE is needed. 

\section{Regularized Maximum Likelihood Estimation for the MoE model}
\label{sec: RMoE}
Regularized MLE allows the selection of a relevant subset of features for prediction and thus encourages sparse solutions. This approach also bounds the norm of the estimated parameters. Hence, it avoids the singularity of the penalized log-likelihood. In mixture-of-experts modeling, one may consider both sparsity in the feature space of the gates, and of the experts. 
As proposed, the MoE model inferred by maximizing a regularized log-likelihood criterion and encourages sparsity for both the gating network parameters and the experts network parameters. This does not require any approximation along with performing the maximization, therefore avoid matrix inversion. 
The proposed regularization that combines two Lasso penalties for the experts parameters, and for the gating network is defined by:
\begin{equation}
\label{eq:PenLoglik MoE}
PL(\bstheta) = L(\bstheta) - \sum_{k=1}^K\lambda_k\|\bsbeta_k\|_1 - \sum_{k=1}^{K-1}\gamma_k\|\bsw_k\|_1.
\end{equation}where $\|\bsv\|_1 = \sum_{j=1}^p|v_j|$ is the $\ell_1$ norm of a vector $\bsv\in \R^p$, $\lambda_k\ge 0$ for all $k = 1,\dots, K$ and $\gamma_k\ge 0$ for all $k = 1,\dots, (K-1)$. Here, \(\bsbeta_k\) is a subvector of \(\bstheta_k\), the parameter vector for the \(k\)th expert. The regularization parameters $\lambda_k$ and $\gamma_k$ control the amount of shrinkage on the parameters $\bsbeta_k$ and $\bsw_k$.
A similar strategy has been proposed in \cite{Kha10} where the author proposed regularization methods for Gaussian regression based on two well-known penalized techniques: Lasso \citep{Tib96} and SCAD \citep{Fan01} which are then approximated in the EM algorithm of the model inference. An $\ell_2$ penalty function for the gating network is added to avoid wildly large positive and negative estimates of the regression coefficients corresponding to the mixing proportions. This behavior can be observed in multinomial regression when the number of potential features is large and they are highly correlated \citep{Park2007b, Bunea2008}.
However, when the correlations are not strong, the effect of the quadratic penalty with a small $\ell_2$ is negligible. We therefore remove this $\ell_2$ penalty from the proposed model. For parameter estimation, Khalili introduced an EM algorithm follows the suggestion of \cite{Hun05} to approximate the penalty function in a some neighborhood by a local quadratic function. After that, a Newton-Raphson can be used to update parameters in the M-step.  
To avoid this numerical instability of the algorithm due to the small values of some of the features in the denominator of this approximation, \cite{Kha10} replaced that  approximation by an $\epsilon$-local quadratic function. 
Unfortunately, these strategies have some drawbacks. First, by approximating the penalty functions with $\epsilon$-quadratic functions, none of the components will be exactly zero. Hence, a threshold should be considered to declare a coefficient as zero, and this threshold affects the degree of sparsity. Secondly, using Newton-Raphson procedure for maximizing a concave function with large dimension $p$ is not an appropriate choice related to the required hessian matrix inversion.\vk 
In a similar scenario, \cite{Per14} suggested an EM algorithm for the regularized MoE of multinomial regression, in which using a transformation that implies inverting the soft-max function. However, there is no evidence to ensure the increasing of their penalized log-likelihood values and this leads to the poor results from their approach. Recently, \cite{Cham18} suggested another approach to the estimation and feature selection in MoE by using an EM algorithm with coordinate ascent updates to overcome these limitations of Khalili's method. 
But this proposal still has some drawbacks since it requires significant computing time due to the maximization of nonsmooth univariate concave function using the Newton method. Hence, it is needed to be improved to deal with large scale data sets. In our approach presented here, we propose an EM algorithm which relies on proximal Newton-type procedures in the M-step to overcome these limitations. We consider that in mixture of experts with three different models for the experts, that is Gaussian, Poisson, and multinomial regressors. 
 
\subsection{Parameter estimation with a proximal Newton-EM algorithm}
\label{ssec: EM}
For each of the three considered GLM for the MoE models, we propose an EM algorithm to monotonically find at least local maximizers of (\ref{eq:PenLoglik MoE}).
The E-step is common to the three models. For the M-step, two different algorithms are proposed to update the model parameters. 
 Specifically, the first one relies on proximal Newton method, while the second one uses a proximal Newton-type method to update the gating network and expert's parameters. The difference between these algorithms is that the proximal Newton-type method we construct here to update the gating network can avoid the numerical instability of the proximal Newton method due to the small value of the mixing proportions. We discuss this difference in Section \ref{sec:GatingNetworkUpdating}.
The EM algorithm for the maximization of (\ref{eq:PenLoglik MoE}) requires the construction of  the penalized complete-data log-likelihood, which is, in our context, given by
{\begin{equation}
PL_c(\bstheta) = L_c(\bstheta)- \sum_{k=1}^K\lambda_k\|\bsbeta_k\|_1 - \sum_{k=1}^{K-1}\gamma_k\|\bsw_k\|_1
\label{eq:complete log-lik RMoE}
\end{equation}}where
\begin{equation*}
L_c(\bstheta) = \sum_{i=1}^{n}\sum_{k=1}^{K} Z_{ik} \log \left[\pi_k(\bsx_i;\bw) p_k(y_i|\bsx_i;\bstheta_k)\right]
\end{equation*}
is the standard complete-data log-likelihood for the MoE model where $Z_{ik}$ an indicator binary-valued variable such that $Z_{ik}=1$ if $Z_i=k$ (i.e., if the $i$th pair $(\bsx_i,y_i)$ is generated from the $k$th expert component) and $Z_{ik}=0$ otherwise. 
Thus, the proposed EM algorithm for the regularized MoE model in its general form runs as follows. After starting with an initial solution $\bstheta^{[0]}$, it alternates between the two following steps until convergence (e.g., when there is no longer a significant change in the relative variation of (\ref{eq:PenLoglik MoE})).
\paragraph{E-step:}
\label{ssec: E-step RMoE} The E-Step computes the conditional expectation of the penalized complete-data log-likelihood (\ref{eq:complete log-lik RMoE}),  given the observed data $\cD$ and a current parameter vector $\bstheta^{[q]}$, $q$ being the current iteration number of the block-wise EM algorithm: 
{\begin{align}
Q(\bstheta;\bstheta^{[q]}) &=  \E\left[PL_c(\bstheta)|\cD;\bstheta^{[q]}\right] \nonumber\\
&= \sum_{i=1}^{n}\sum_{k=1}^{K}\tau_{ik}^{[q]} \log \left[\pi_k(\bsx_i;\bw) p_k(y_{i}|\bsx_i;\bstheta_{k})\right]
- \sum_{k=1}^K\lambda_k\|\bsbeta_k\|_1 - \sum_{k=1}^{K-1}\gamma_k\|\bsw_k\|_1
\label{eq:Q-function RMoE}
\end{align}}where
{\begin{eqnarray}
\tau_{ik}^{[q]} = \Pro(Z_i=k|y_{i},\bsx_i;\bstheta^{[q]})
= \pi_k(\bsx_i;\bw^{[q]})p_k(y_{i}|\bsx_i;\bstheta_{k}^{[q]})/p(y_{i}|\bsx_i;\bstheta^{[q]})
\label{eq:RMoE post prob}
\end{eqnarray}}is the conditional probability that the data pair $(\bsx_i,y_i)$ is generated by  the $k$th expert.
This step only requires  the computation of the conditional component probabilities $\tau^{[q]}_{ik}$ $(i=1,\ldots,n)$ for each of the $K$ experts. 
\paragraph{M-step:} The M-Step updates the parameters by maximizing the $Q$ function (\ref{eq:Q-function RMoE}) w.r.t $\bstheta$. The Q-function can be written as:
\begin{equation*}
Q(\bstheta;\bstheta^{[q]})  = Q(\bw;\bstheta^{[q]}) + \sum\limits_{k=1}^KQ_k(\bstheta_k;\bstheta^{[q]})
\end{equation*}
with
{\begin{align}
 Q(\bw;\bstheta^{[q]}) &= \sum_{i=1}^n\sum_{k=1}^K\tau_{ik}^{[q]}\log\pi_k(\bsx_i;\bw)-  \sum_{k=1}^{K-1}\gamma_k\|\bsw_k\|_1,\notag\\
				    &= \underbrace{\sum_{i=1}^n\sum_{k=1}^{K-1}\tau_{ik}^{[q]}(w_{k0}+\bsx_i^\top \bsw_k)- \sum_{i=1}^n\log\Bigl[1+\sum_{k=1}^{K-1}e^{w_{k0}+\bsx_i^\top \bsw_k}\Bigl]}_{I(\bw)} - \sum_{k=1}^{K-1}\gamma_k\|\bsw_k\|_1.\label{QP}
\end{align}}
and
{\begin{equation*}
Q_k(\bstheta_k;\bstheta^{[q]}) =  \sum_{i=1}^n\tau_{ik}^{[q]}\log p_k(y_{i}|\bsx_i;\bstheta_{k}) - \lambda_k\|\bsbeta_k\|_1.
\end{equation*}}The parameters $\bw$ are therefore updated by maximizing the function (\ref{QP}).
Here, the composite function $Q(\bw;\bstheta^{[q]})$ is concave and does not have the  weighted Lasso form. One can use coordinate ascent algorithm to update $\bw$ since the penalty part has a separate structure (see \cite{Tse01} for more details). However, this approach requires a lot of computing and is not suitable for large scale data (see \cite{Cham18}). In this case, proximal Newton algorithm and proximal Newton-type algorithm are good choices to overcome these drawbacks. The principle of these methods are described in Appendix \ref{appendixA}. The idea of these approaches lies in the fact that they approximate the smooth part of $Q(\bw;\bstheta^{[q]})$ with a local quadratic function. After that, one will solve a weighted Lasso regression problem, which has a closed-form update. The solution of this weighted Lasso regression provides a direction that one can choose to improve the value of $Q(\bw;\bstheta^{[q]})$ using backtracking line search.\vk 
The methods for updating the gating network's parameters using proximal Newton, and proximal Newton-type method are described in the next section.
\subsection{Proximal Newton-type procedure for updating the gating network}\label{sec:GatingNetworkUpdating}
In this part, we propose two approaches for updating the gating network parameters $\bw=\{(w_{k0}, \bsw_k)\}$ 
by maximizing $Q(\bw; \bstheta^{[q]})$ based on the proximal Newton and the proximal Newton-type method. More specifically, in this case the algorithms first apply the (block) coordinate ascent algorithm and then apply the proximal Newton-type algorithm for each block problem.

First, partial Newton steps are performed by forming a partial quadratic approximation to $Q(\bw;\bstheta^{[q]})$ (Taylor expansion at the current estimates), allowing only $(w_{k0},\bsw_k)$ to vary for a single class at a time. This algorithm is similar to the one in \cite{Friedman10} except the fact that here after each outer loop that cycles over $k$, a backtracking line search is performed over the step size parameter $t\in[0,1]$. 
The partial quadratic approximation to $I(\bw)$ in (\ref{QP}) w.r.t $(w_{k0}, \bsw_k)$ at $\tilde{\bw}$ is given by (see Appendix  \ref{appendixB} for more details)
\begin{equation*}
l_{I_k}(w_{k0},\bsw_k) = -\frac{1}{2}\sum\limits_{i=1}^nd_{ik}(c_{ik} - w_{k0} - \bsx_i^\top\bsw_k)^2 + C(\tilde{\bw}),
\end{equation*}
where
\begin{align}
c_{ik} & = \tilde{w}_{k0} + \bsx_i^\top\tilde{\bsw}_k + \frac{\tau_{ik}^{[q]} - \pi_k(\tilde{\bw};\bsx_i)}{\pi_k(\tilde{\bw};\bsx_i)(1-\pi_k(\tilde{\bw};\bsx_i))},\label{eq:cik}\\
d_{ik} & = \pi_k(\tilde{\bw};\bsx_i)(1-\pi_k(\tilde{\bw};\bsx_i)),\label{eq:dik}
\end{align}
and $ C(\tilde{\bw})$ is a function of $\tilde{\bw}$. After calculating the partial quadratic approximation $l_{I_k}(w_{k0},\bsw_k)$ about the current parameters $\tilde{\bw}$, a coordinate ascent algorithm is used to solve the penalized weighted least-square problem
\begin{equation}\label{weightedLS}
\max\limits_{(w_{k0},\bsw_k)} l_{I_k}(w_{k0},\bsw_k) - \gamma_k\|\bsw_k\|_1.
\end{equation}
Using the soft-thresholding operator (see \cite[sec. 5.4]{TH15c}), one can obtain the closed-form update for $w_{kj}$ as follows
\begin{equation}\label{Update:wkj}
w_{kj}^{m+1} = \frac{\mathcal{S}_{\gamma_k}(\sum\limits_{i=1}^nd_{ik}u_{ikj}^mx_{ij})}{\sum\limits_{i=1}^nd_{ik}x_{ij}^2},
\end{equation}
with $u_{ikj}^m = c_{ik}-w_{k0}^m - \bsx_i^\top\bsw_k^m + w_{kj}^mx_{ij}$ and $\mathcal{S}_{\gamma_k}(.)$ is a soft-thresholding operator defined by $[{\bf\mathcal{S}}_\gamma(u)]_j = \text{sign}(u_j)(|u_j|-\gamma)_+$ and {$(x)_+$}  a shorthand for $\max\{x, 0\}$. Here, $m$ is defined as the $m$th step of the coordinate ascent algorithm. Note that, for each iteration of the coordinate ascent algorithm one parameter is updated while other are kept fixed, that means for $h\not=j$, $w_{kh}^{m+1} = w_{kh}^m$. For $w_{k0}$, the closed-form update is given by
\begin{equation}\label{Update:wk0}
w_{k0}^{m+1} = \frac{\sum\limits_{i=1}^nd_{ik}(c_{ik}-\bsx_i^\top\bsw_k^m)}{\sum\limits_{i=1}^nd_{ik}}.
\end{equation}
Finally, once the coordinate ascent algorithm converges, the new values of $(w_{k0},\bsw_k)$ are taken into account for the next loop of the proximal Newton algorithm. Overall, the algorithm is summarized by pseudo-code \ref{algo1}.
\begin{algorithm}\label{algo:PN}
\caption{Proximal Newton method for updating the gating network}
\label{algo1}
\begin{algorithmic}[1]
\STATE Initialize $\mathbf{w}^{(0)} \gets \mathbf{w}^{[q]}$, $s \gets 0$
\REPEAT
    \FOR{$k = 1$ to $K-1$}
        \STATE Compute $c_{ik}$ and $d_{ik}$ via \eqref{eq:cik}--\eqref{eq:dik} with $\tilde{\mathbf{w}} = \mathbf{w}^{(s)}$
        \STATE Solve penalized weighted least-squares \eqref{weightedLS} via coordinate ascent:
        \STATE \quad Update intercept $w_{k0}$ via \eqref{Update:wk0}
        \STATE \quad For $j = 1,\dots,p$: update $w_{kj}$ via \eqref{Update:wkj} with soft-thresholding $\mathcal{S}_{\gamma_k}$
        \STATE Denote solution as $\tilde{\mathbf{w}}_k^{(s)}$
        \STATE Set $(w_{k0}^{(s)}, \boldsymbol{w}_k^{(s)}) \gets \tilde{\mathbf{w}}_k^{(s)}$
    \ENDFOR
    \STATE Perform backtracking line search for $t \in (0,1]$ satisfying:
    \STATE \quad $Q\bigl((1-t)\mathbf{w}^{(s)} + t\tilde{\mathbf{w}}^{(s)}; \bstheta^{[q]}\bigr) \geq Q\bigl(\mathbf{w}^{(s)}; \bstheta^{[q]}\bigr)$
    \STATE Update: $\mathbf{w}^{(s+1)} \gets (1-t)\mathbf{w}^{(s)} + t\tilde{\mathbf{w}}^{(s)}$
    \STATE $s \gets s + 1$
\UNTIL the stopping criterion is satisfied.
\end{algorithmic}
\end{algorithm}\\
In this EM algorithm, the backtracking line-search is needed for the algorithm to converge to the optimal solution. The proximal Newton method presented here can overcome the drawback of the coordinate ascent algorithm in \cite{Cham18} since at each step there is a closed-form update for each parameter. Hence, it improves the run time of the algorithm. \vk
\medskip
Even though in some cases the values of the probabilities $\pi_k(\tilde{\bw};\bsx_i)$ can become too small (or too close to $1$), and the algorithm can get stuck while solving (\ref{weightedLS}). 
To address this issue, we consider the proximal Newton-type method as a proper choice for this situation. Proximal Newton-type methods use a symmetric negative definite matrix ${\bf B}\approx \triangledown^2 I(\tilde{\bw}_k)$ to model the curvature of $I(\bw)$ at $(w_{k0},\bsw_k)$. In this case, one can follow the suggestions of \cite[sec. 8.7]{Lan13} and \cite{Gormley2008} by choosing a constant negative definite matrix $\bf{B}$ such as $\triangledown^2 I(\tilde{\bw}_k) > {\bf B}$. The proximal Newton-type algorithm  here can be interpreted as a special case of the MM algorithm \citep{Hun04}. 
Specifically it is a  minorize-maximize algorithm for updating the gating network and also the expert network in multinomial outputs case.\vk
Since,
$$\frac{\partial^2 I(\bw)}{\partial w_{kj}\partial w_{kh}} = -\sum\limits_{i=1}^n x_{ij}x_{ih}\pi_k(\bsx_i;\bw)(1-\pi_k(\bsx_i;\bw)),\ \forall j, h,$$
then, using the fact that $\pi(1-\pi) \le 1/4$, we can take ${\bf B} = -1/4\sum_{i=1}^n \bsx_i\bsx_i^\top$. Thus, instead of solving (\ref{weightedLS}), one can solve the local quadratic model
\begin{equation*}
\max\limits_{(w_{k0},\bsw_k)} \hat{l}_{I_k}(w_{k0},\bsw_k) - \gamma_k\|\bsw_k\|_1.
\end{equation*}
where
\begin{equation*}
\hat{l}_{I_k}(w_{k0},\bsw_k) = -\frac{1}{8}\sum\limits_{i=1}^n(\hat{c}_{ik} - w_{k0} - \bsx_i^\top\bsw_k)^2 +\hat{C}(\tilde{\bw}),
\end{equation*}
and
\begin{equation*}
\hat{c}_{ik} = \tilde{w}_{k0} + \bsx_i^\top\tilde{\bsw}_k + 4(\tau_{ik}^{[q]} - \pi_k(\tilde{\bw};\bsx_i)),
\end{equation*}
$\hat{C}(\tilde{\bw})$ is a function of $\tilde{\bw}$. Here, it is clear that this approach has some advantages. One can avoid computing the Hessian matrix and can also avoid numerical instability caused by $\pi_k(\tilde{\bw};\bsx_i)$. In addition, the backtracking line search procedure is also not necessary when using this approach. The increase of the $Q(\bw; \bstheta^{[q]})$ after each loop is guaranteed, since this algorithm is a proximal Newton-type algorithm and is a specific case of the MM algorithm. 

\subsection{Updating the Experts Network}
\label{sec:updating_experts}

We now address the update of the expert network parameters $\bm{\theta}_k$ for $k = 1, \ldots, K$. A key advantage of our approach is that the proximal Newton strategy can be applied uniformly when the expert models belong to the generalized linear model (GLM) family. We consider experts whose conditional density follows a general exponential family distribution. This provides a unified framework for the Gaussian (continuous), Poisson (count), and Multinomial (categorical) outputs previously discussed.

The conditional density of the response $Y_i$ given the input $\bm{x}_i$ and the expert assignment $Z_i = k$ can be expressed in the canonical form:

\begin{equation*}
p_k(y_i | \bm{x}_i; \bstheta_k) = \exp \left\{ \frac{y_i \eta_k(\bm{x}_i) - b(\eta_k(\bm{x}_i))}{a(\phi_k)} + c(y_i, \phi_k) \right\},
\end{equation*}

where \(\eta_k(\boldsymbol{x}_i) = \beta_{k0} + \bm{x}_i^\top \bsbeta_k \) is the natural parameter and linear predictor, $b(\cdot)$ is the log-partition function,
 $a(\phi_k)$ is the dispersion parameter, and $c(\cdot, \cdot)$ is the base measure. Here, \(\boldsymbol{\theta}_k = (\beta_{k0}, \boldsymbol{\beta}_k^\top, \phi_k)^\top\) denotes the complete parameter vector for the $k$-th expert.

The objective for the $k$-th expert in the M-step is to maximize the penalized Q-function:

\begin{equation*}
Q_k(\bstheta_k; \bstheta^{[q]}) = \sum_{i=1}^n \tau_{ik}^{[q]} \log p_k(y_i | \bm{x}_i; \bstheta_k) - \lambda_k \|\bm{\beta}_k\|_1.
\end{equation*}

Substituting the exponential family form, we obtain:

\begin{equation}
Q_k(\bm{\theta}_k; \bm{\theta}^{[q]}) = \underbrace{\sum_{i=1}^n \tau_{ik}^{[q]} \left[ \frac{y_i \eta_k(\bm{x}_i) - b(\eta_k(\bm{x}_i))}{a(\phi_k)} \right]}_{P_k(\bstheta_k; \bm{\theta}^{[q]})} - \lambda_k \|\bm{\beta}_k\|_1 + \text{constant}.
\label{eq:Q_expert_decomposed}
\end{equation}

For canonical link functions, the function $P_k(\bstheta_k; \bm{\theta}^{[q]})$ is concave in $\bm{\theta}_k$, making the proximal Newton method applicable. We construct a quadratic approximation to $P_k$ around the current parameter estimate $\tilde{\bstheta}_k$.

The second-order Taylor expansion yields the local quadratic model:

\begin{equation}
\tilde{P}_k(\bstheta_k; \bm{\theta}^{[q]}) \approx P_k(\tilde{\bstheta}_k; \bm{\theta}^{[q]}) + \nabla P_k(\tilde{\bm{\theta}}_k; \bm{\theta}^{[q]})^\top (\bm{\theta}_k - \tilde{\bm{\theta}}_k) + \frac{1}{2} (\bm{\theta}_k - \tilde{\bm{\theta}}_k)^\top \nabla^2 P_k(\tilde{\bm{\theta}}_k; \bm{\theta}^{[q]}) (\bm{\theta}_k - \tilde{\bm{\theta}}_k).
\label{eq:taylor_expansion}
\end{equation}

It can be shown that this approximation simplifies to a weighted least squares form (see Appendix \ref{app:quadratic_approx} for more details):

\begin{equation}
\tilde{P}_k(\bm{\theta}_k; \bm{\theta}^{[q]}) = -\frac{1}{2} \sum_{i=1}^n \omega_{ik} \left( y_{ik}^* - \beta_{k0} - \bm{x}_i^\top \bm{\beta}_k \right)^2 + C(\tilde{\bm{\theta}}_k),
\label{eq:quadratic_approx_expert}
\end{equation}

where $C(\tilde{\bm{\theta}}_k)$ is a constant depending on $\tilde{\bm{\theta}}_k$, and the {working response} $y_{ik}^*$ and {weights} $\omega_{ik}$ are defined as:

\begin{align*}
y_{ik}^* &= \tilde{\eta}_k(\bm{x}_i) + \frac{y_i - b'(\tilde{\eta}_k(\bm{x}_i))}{b''(\tilde{\eta}_k(\bm{x}_i))}, \\
\omega_{ik} &= \frac{\tau_{ik}^{[q]} }{a(\phi_k)} b''(\tilde{\eta}_k(\bm{x}_i)).
\end{align*}

Here, $b'(\eta)$ and $b''(\eta)$ are the first and second derivatives of the log-partition function, corresponding to the model's mean and variance functions, respectively. The term $\tilde{\eta}_k(\bm{x}_i) = \tilde{\beta}_{k0} + \bm{x}_i^\top \tilde{\bm{\beta}}_k$ represents the current estimate of the linear predictor for expert $k$ at observation $i$, evaluated at the current parameter values $\tilde{\bm{\theta}}_k$.

The M-step update for the expert parameters is then found by solving the penalized weighted least squares problem:

\begin{equation}
\max_{(\beta_{k0}, \bm{\beta}_k)} \left\{ -\frac{1}{2} \sum_{i=1}^n \omega_{ik} \left( y_{ik}^* - \beta_{k0} - \bm{x}_i^\top \bm{\beta}_k \right)^2 - \lambda_k \|\bm{\beta}_k\|_1 \right\}.
\label{eq:expert_lasso_problem}
\end{equation}

This problem is efficiently solved using a coordinate ascent algorithm with soft-thresholding, following a procedure similar to that described in Algorithm \ref{algo1}. The specific forms of $y_{ik}^*$ and $\omega_{ik}$ for the Gaussian, Poisson, and Multinomial models are derived in Appendix~\ref{app:expert_forms}.

In all cases, the proximal Newton-type method can be employed by replacing the Hessian with a suitable global lower bound (e.g., $\mathbf{B} = -\frac{1}{4} \sum_{i=1}^n \tau_{ik}^{[q]} \bm{x}_i \bm{x}_i^\top$) to ensure numerical stability when the weights $\omega_{ik}$ become very small.

\subsection{Model selection and hyperparameter tuning}
In practice, appropriate values for the tuning parameters $(\lambda_k, \gamma_k)$ must be selected. To this end, we adopt a modified BIC combined with a grid-search scheme, extending the criterion used in \cite{Sta10} for regularized mixtures of regressions. Although the tuning parameters are defined component-wise, in our experiments we use common tuning parameters across all components, i.e., $\lambda_k \equiv \lambda$ for $k=1,\ldots,K$ and $\gamma_k \equiv \gamma$ for $k=1,\ldots,K-1$. Thus, for each candidate number of expert components $K \in \{K_1, \ldots, K_M\}$, a grid of common tuning parameters is specified. Let $\lambda$ and $\gamma$ denote the common tuning parameters for the expert and gating networks, respectively. Consider the grids ${\lambda_{(1)}, \ldots, \lambda_{(M_1)}}$ for $\lambda$ and ${\gamma_{(1)}, \ldots, \gamma_{(M_2)}}$ for $\gamma$. The optimal combination $(K,\lambda,\gamma)$ is then selected according to the modified BIC over all candidate values in the corresponding grids.
For each triplet $(K, \lambda, \gamma)$, the maximizer of the penalized log-likelihood, denoted by $\widehat{\bstheta}_{K, \lambda, \gamma}$, is obtained using one of the hybrid EM algorithms described above. The modified BIC criterion is then computed as
\begin{equation*}
\mathrm{BIC}(K, \lambda, \gamma)
= L\big(\widehat{\bstheta}_{K, \lambda, \gamma}\big)
- DF(\lambda, \gamma) \frac{\log n}{2},
\end{equation*}
where $L(\cdot)$ denotes the unpenalized log-likelihood evaluated at the penalized estimator, $n$ is the sample size, $DF(\lambda, \gamma)$ denotes the estimated number of non-zero coefficients in the model. Finally, the model corresponding to $(K, \lambda, \gamma) = (\tilde{K}, \tilde{\lambda}, \tilde{\gamma})$, which maximizes the modified BIC, is selected.\\
Although selecting optimal tuning parameters for penalized MoE models remains an open research problem, the modified BIC demonstrates satisfactory performance in our experiments.  

\section{Experimental study}
\label{sec: Experiments}
The performance of these methods is studied on both simulated data and real data. The results of these algorithms are compared to the standard non-penalized MoE  for the Gaussian model and Poisson model (denoted by GMoE and PMoE respectively) and MoE with $\ell_2$ regularization (MMoE+$\ell_2$) for the multinomial model since local maximum parameters that closed to the true value for the MoE of multinomial model cannot be found in some cases. Several evaluation criteria are used to assess the performance of the models, including sparsity, parameters estimation and clustering criteria. Our penalized models are denoted by GMoE-Lasso, PMoE-Lasso and MMoE-Lasso for the Gaussian, Poisson and multinomial cases, respectively.\vk
The R packages of codes of the developed algorithms and the documentation are publicly available on Github via this link\footnote{\url{https://github.com/nv-thin/GLM-RMoE}}.

\subsection{Evaluation criteria}

The results of all the models are compared based on three different criteria: sensitivity/specificity, parameter estimation, and clustering performance on the
simulated data. For the first criterion, we follow the terminology of \cite{Kha10}. We emphasize that this usage differs from the conventional one in classification, where sensitivity refers to the detection of nonzero (signal) coefficients; here the roles are reversed. Concretely, we define
\begin{itemize}
\item {\it Sensitivity} (true-zero rate): the proportion of truly zero coefficients that are correctly estimated as zero;
\item {\it Specificity} (true-nonzero rate): the proportion of truly nonzero coefficients that are correctly estimated as nonzero.
\end{itemize}
Thus, sensitivity measures the ability of the method to shrink irrelevant coefficients exactly to zero, while specificity measures its ability to retain the
relevant ones. 

In this way, the ratio of the estimated zero/nonzero coefficients to the true number of zero/nonzero coefficients of the true parameter is computed for each component. In our simulation, the proportion of correctly estimated zero coefficients and nonzero coefficients have been calculated for each data set  for the experts parameters and the gating parameters. We present the average proportion of these criteria computed over 100 different data sets. To deal with the label switching before calculating these criteria, we permuted the estimated coefficients based on an ordered between the expert parameters. If the label switching happens, one can permute the expert parameters and the gating parameters then replace the $k$th gating network vector with $\bw_k^{per}=\bw_k - \bw_K$. By doing so, we ensure that the log-likelihood will not change, that means $L(\hat{\bstheta}) = L(\hat{\bstheta}^{per})$ and these parameters satisfy the initialized condition $\bw_K^{per} = \boldsymbol{0}$. However, the penalized log-likelihood value can be different from the one before permutation. So this may result in misleading values of the sparsity criterion of the model when we permute the parameters. The regularized  method tends to choose the model with small absolute values of the gating network. Fortunately, for $K = 2$, the log-likelihood function and the penalized log-likelihood function will not change since we have $\bw_1^{per} = -\bw_1$.\\
For the second criterion of parameter estimation, we compute the mean and standard deviation for both the penalized parameters and the non penalized parameters and compare with the true value $\bstheta$. We also consider the mean squared error (MSE) between each component of the true parameter vector and the estimated one, which is given by $\|\theta_j - \hat\theta_j\|_2^2$. \\
For the clustering criterion, once the parameters are estimated and permuted, the provided conditional component probabilities $\hat{\tau}_{ik}$ defined in (\ref{eq:RMoE post prob}) represent a soft partition of the data. A hard partition of the data is given by applying the Bayes's allocation rule
$$\hat{z}_i = \arg\max_{k =1}^K \tau_{ik}(\widehat{\bstheta}),$$ 
where $\hat{z}_i$ represents the estimated cluster label for the $i$th observation. 
Given the estimated and true cluster labels, we compute the  correct classification rate and the Adjusted Rand Index (ARI), which lies in $[-1,1]$; it equals 1 for perfect agreement, has expected value 0 under random labeling, and can be negative if agreement is worse than chance.

\subsection{Simulation study}
For each data set, consider $n=300$ observations $\bsx \in \mathbb{R}^{p}$ with $p=9$, generated from a multivariate Gaussian distribution with zero mean and correlation defined by $\text{corr}(x_{ij}, x_{ij\prime}) = 0.5^{|j-j\prime|}$. The response $Y|\bsx$ is generated from a normal MoE model, a multinomial model with two classes and a Poisson model of $K = 2$ expert components with the following regression coefficients:\\
$\bullet$ Parameters for the normal MoE model:
\begin{align*}
(\beta_{10},\bsbeta_1)^\top &= (0, 0, 1.5, 0, 0, 0, 1, 0, 0, -2)^\top;\\
(\beta_{20},\bsbeta_2)^\top &= (0, 1, -1.5, 0, 0, 2, 0, 0, 0, 0)^\top;\\
(w_{10}, \bsw_1)^\top &=  (1, 2, 0, 0, -1, 0, 1, 0, 0, 0)^\top;\\
\sigma_1 = \sigma_2 &= \sigma = 1.
\end{align*}
$\bullet$ Parameters for the Poisson model:
\begin{align*}
(\beta_{10},\bsbeta_1)^\top &= (0, 1, 0, -2, 0, 1.5, 0, 0, 0, 0)^\top;\\
(\beta_{20},\bsbeta_2)^\top &= (0, 0, 2, 0, -1, 0, 0, 0, 0, 0)^\top;\\
(w_{10}, \bsw_1)^\top &=  (1, 0, 0, 1, 0, -1.5, 0, -1, 0, 0)^\top.
\end{align*}
$\bullet$ Parameters for the multinomial model ($R=2$): 
\begin{align*}
(\beta_{110},\bsbeta_{11})^\top & = (0, -1, 2, 0, 0, 1.5, 0, 0, 0, 0)^\top;\\
(\beta_{210},\bsbeta_{21})^\top & = (0, 1, 0, 0, -2, 0, 0, 0, 0, 0)^\top;\\
(w_{10}, \bsw_1)^\top &=  (1, 0, 0, 1, 0, 0, -1.5, 0, 0, 0)^\top.
\end{align*}
100 data sets were generated for each simulation. The results will be presented in the following sections.


\subsubsection{Sensitivity/specificity criteria}

Table \ref{S/S} presents the sensitivity ($S_1$) and specificity ($S_2$) values for experts $1$ and $2$ and for the gating network, for each of the models considered. The non-penalized MoE models cannot be regarded as model-selection procedures, since their sensitivity (true-zero rate) is effectively zero; they rarely estimate any coefficient exactly as zero; the corresponding results are therefore omitted. In particular, estimation for the multinomial model under the standard (unpenalized) MoE is unstable: for a typical data set, a local maximizer close to the true parameter value could not be found (see Table \ref{Multinomial}). This is consistent with \cite{Rosset2004}, who showed that in multinomial regression the maximum-likelihood solution is undefined when the data are separable by the predictors. To address this, \cite{Park2007a} add a small
quadratic ($\ell_2$) penalty, so that as the Lasso tuning parameter $\lambda$
approaches zero the coefficients converge to the $\ell_2$-penalized multinomial solution rather than diverging to infinity. Based on this idea, we add two small quadratic penalizations to the MoE model with multinomial outputs, 

$$\frac{\rho}{2}\sum\limits_{k=1}^K\sum\limits_{r=1}^{R-1}\|\bsbeta_{kr}\|_2^2 \text{ and } \frac{\rho}{2}\sum\limits_{k=1}^{K-1}\|\bsw_k\|_2^2,$$
where $\rho = 0.1\log(n)$ as in \cite{Kha10}. The Lasso-regularized MoE in the multinomial case is then compared with this $\ell_2$-regularized MoE. The Lasso performs well in detecting nonzero coefficients in both the experts and the gating network, and the penalty removes the instability of the estimators. When the features are strongly correlated, adding an $\ell_2$ penalty to the experts and the gating network may further improve stability; we do not pursue this variant here.

\begin{table}[!h]
\centering
\begin{tabular}{|c|c|c|c|c|c|c|}
\hline
Model &\multicolumn{2}{c|}{Expert 1} &\multicolumn{2}{c|}{Expert 2}& \multicolumn{2}{c|}{Gate}\\
\cline{2-7}
  & $S_1$ & $S_2$ & $S_1$ & $S_2$ & $S_1$ & $S_2$\\
\hline
GMoE-Lasso & 0.666 & 1.000 & 0.767 & 0.983 & 0.950 & 0.953 \\ 
PMoE-Lasso & 0.696 & 1.000 & 0.752 & 1.000 & 0.830 & 0.998 \\ 
MMoE-Lasso & 0.566 & 0.923 & 0.703 & 0.705 & 0.706 & 0.940 \\ 
\hline
\end{tabular}
\caption{Sensitivity ($S_1$) and specificity ($S_2$) results of the Lasso regularized MoE models for Gaussian outputs (GMoE-Lasso), 
Poisson outputs (PMoE-Lasso) and multinomial outputs (MMoE-Lasso). \label{S/S}}
\end{table}
\begin{table}[!h]
\centering
\begin{tabular}{|c|c|c||c|c|c||c|c|c|}
\hline
\multicolumn{3}{|c||}{True value} & \multicolumn{3}{c||}{MMoE-Lasso} & \multicolumn{3}{c|}{MoE}\\
\hline
Exp. 1 & Exp. 2 & Gate & Exp. 1 & Exp. 2 & Gate & Exp. 1 & Exp. 2 & Gate \\
\hline
0 & 0 & 1 & 0.1973 & -0.5072 & 0.9429 & 2.6323 & \textbf{-38.5145} & 0.4642 \\ 
  -1 & 1 & 0 & -0.7885 & 0.4322 & 0 & -1.6220 & \textbf{9.8878} & -0.2178 \\ 
  2 & 0 & 0 & 1.3598 & 0 & 0 & 3.9504 & \textbf{-6.3021} & 0.4919 \\ 
  0 & 0 & 1 & 0 & 0 & 0.8351 & 0.6870 & \textbf{10.3141} & -0.1080 \\ 
  0 & -2 & 0 & 0 & -1.0807 & 0 & -5.0639 & \textbf{13.3031} & 0.3521 \\ 
  1.5 & 0 & 0 & 1.3942 & 0 & 0 & 4.2943 & \textbf{-11.2733} & 0.1669 \\ 
  0 & 0 & -1.5 & 0 & 0 & -1.6670 & -0.4807 & \textbf{17.2681} & -0.3019 \\ 
  0 & 0 & 0 & 0 & 0 & 0 & -1.1754 & \textbf{13.3834} & 0.0193 \\ 
  0 & 0 & 0 & 0 & 0 & -0.0103 & -1.4630 & \textbf{26.3915} & 0.3007 \\ 
  0 & 0 & 0 & 0.1507 & 0 & 0 & 3.3752 & \textbf{-26.3064} & -0.2485 \\ 
\hline
\end{tabular}
\caption{Estimated parameters for a typical data set obtained with Lasso regularized MoE (MMoE-Lasso) and MLE for MoE with multinomial observations. \label{Multinomial}}
\end{table}
\subsubsection{Parameter estimation}
\begin{figure*}[!h]
\centering
\begin{tabular}{ccc}
\includegraphics[width = 5 cm, height = 5 cm]{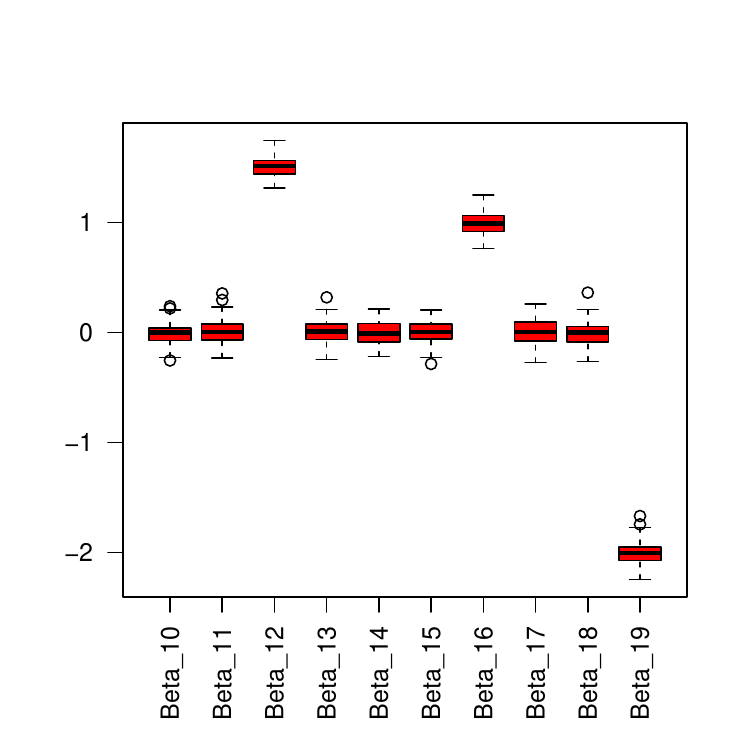}&
\includegraphics[width = 5 cm, height = 5 cm]{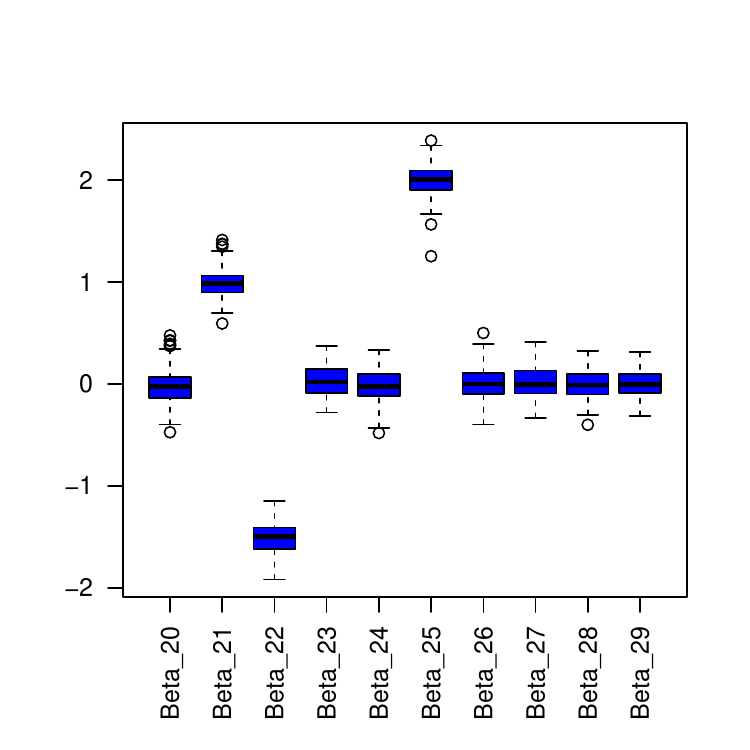}&
\includegraphics[width = 5 cm, height = 5 cm]{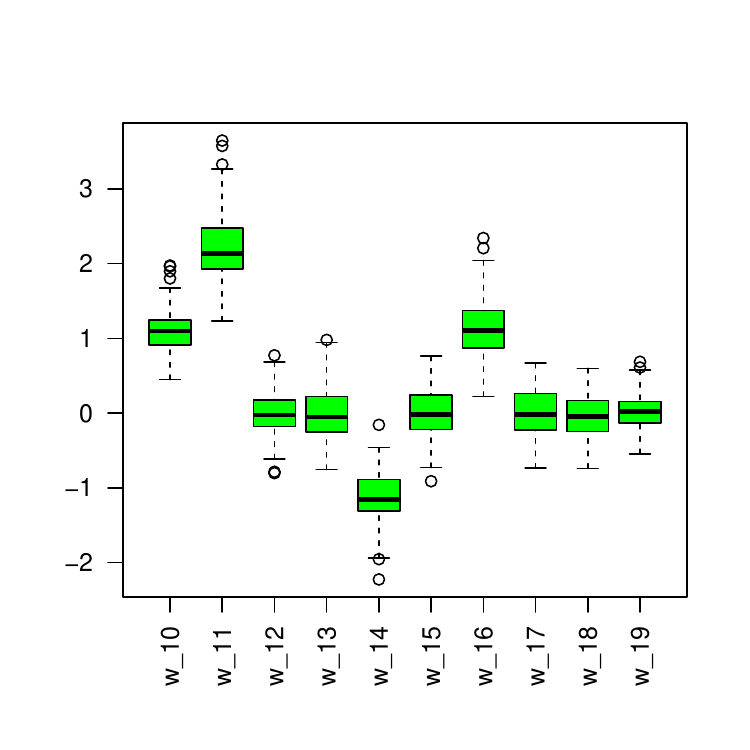}\\
GMoE-Exp.1 & GMoE-Exp.2 & GMoE-Gate\\
\includegraphics[width = 5 cm, height = 5 cm]{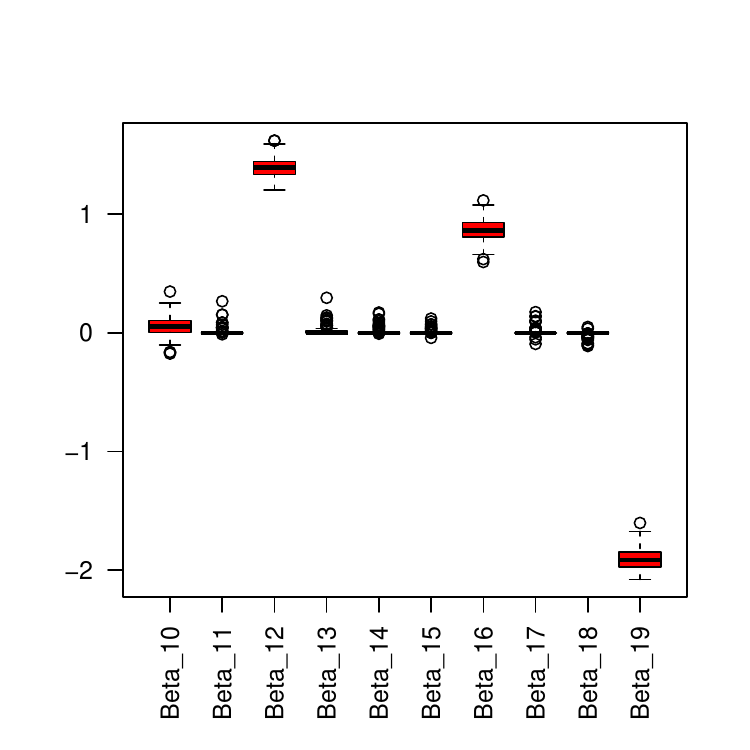}&
\includegraphics[width = 5 cm, height = 5 cm]{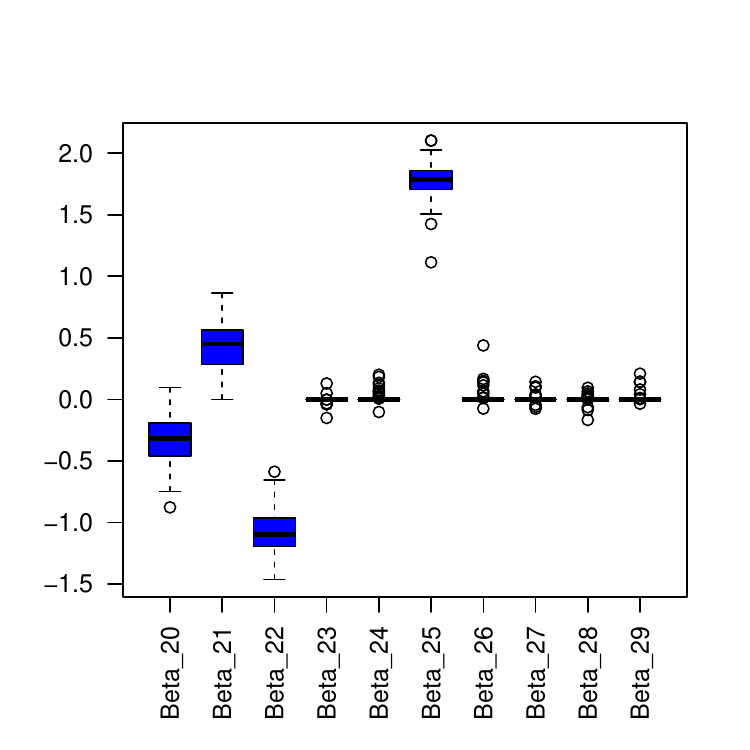}&
\includegraphics[width = 5 cm, height = 5 cm]{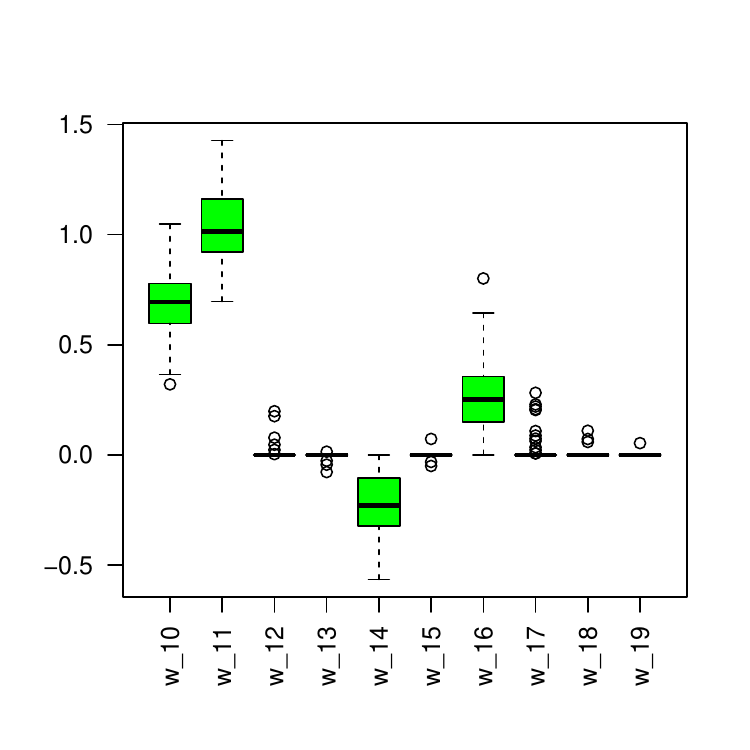}\\
GMoE-Lasso-Exp.1 & GMoE-Lasso-Exp.2 & GMoE-Lasso-Gate
\end{tabular}
\caption{Boxplots of non-penalized MoE model (GMoE) and the proposed Lasso penalized MoE model (GMoE-Lasso) for Gaussian case.\label{Fig:Gaussian}}
\end{figure*}

\begin{figure*}[!h]
\centering
\begin{tabular}{ccc}
\includegraphics[width = 5 cm, height = 5 cm]{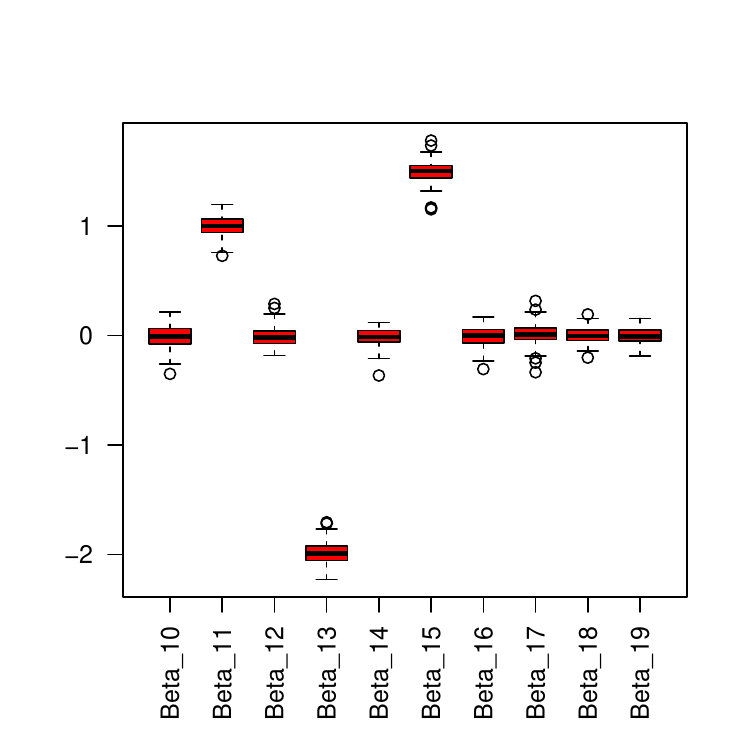}&
\includegraphics[width = 5 cm, height = 5 cm]{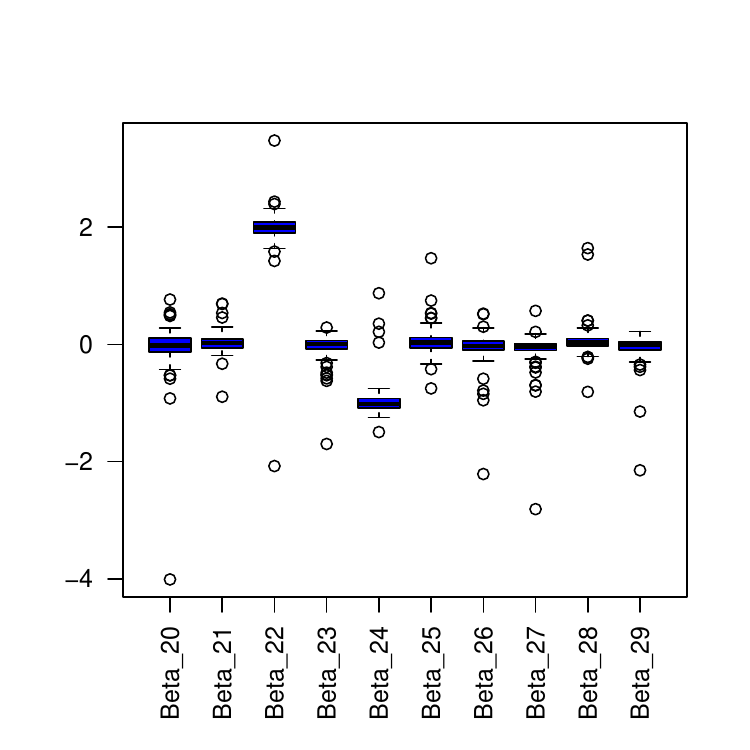}&
\includegraphics[width = 5 cm, height = 5 cm]{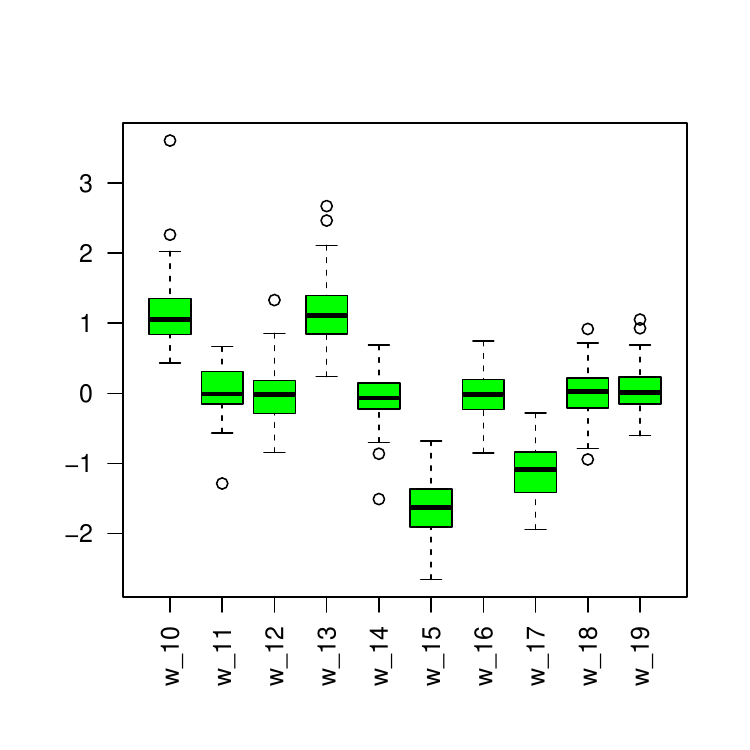}\\
PMoE-Exp.1 & PMoE-Exp.2 & PMoE-Gate\\
\includegraphics[width = 5 cm, height = 5 cm]{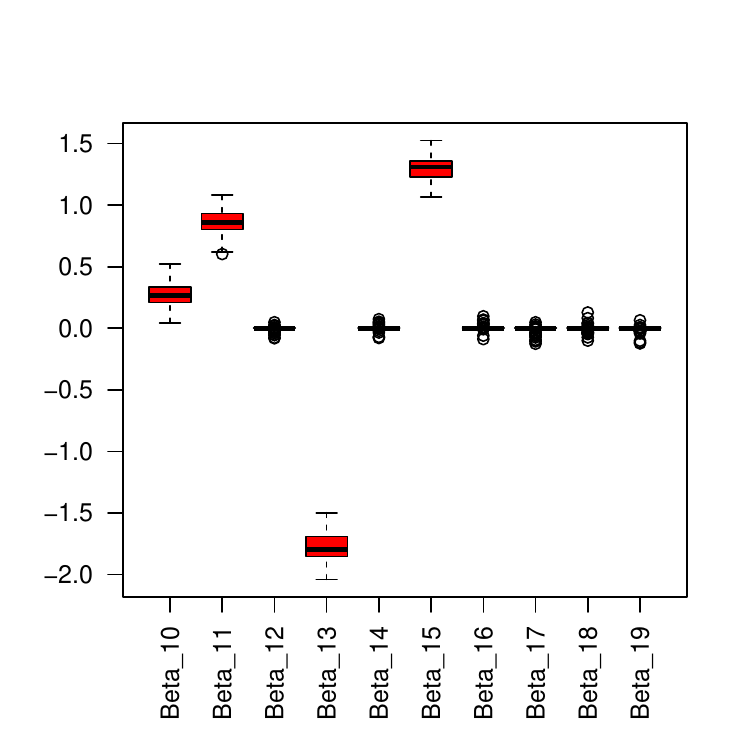}&
\includegraphics[width = 5 cm, height = 5 cm]{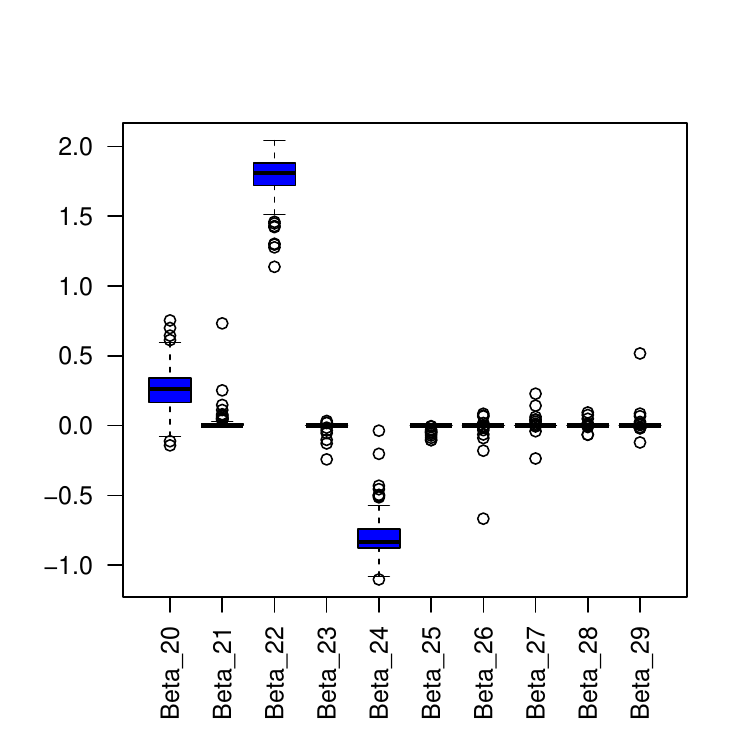}&
\includegraphics[width = 5 cm, height = 5 cm]{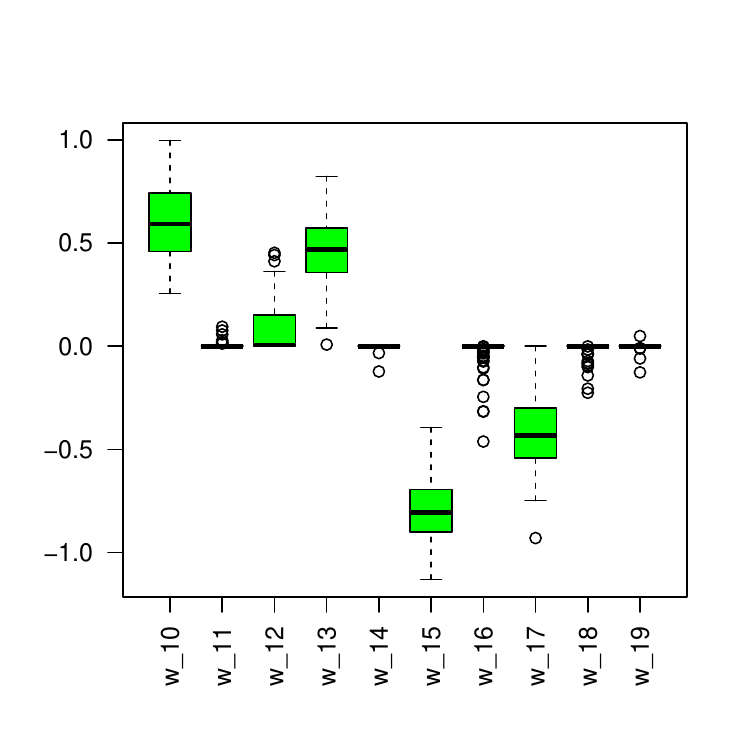}\\
PMoE-Lasso-Exp.1 & PMoE-Lasso-Exp.2 & PMoE-Lasso-Gate
\end{tabular}
\caption{Boxplots of non-penalized MoE model (PMoE) and the proposed Lasso penalized MoE model (PMoE-Lasso) for Poisson case.\label{Fig:Poisson}}
\end{figure*}

\begin{figure*}[!h]
\centering
\begin{tabular}{ccc}
\includegraphics[width = 5 cm, height = 5 cm]{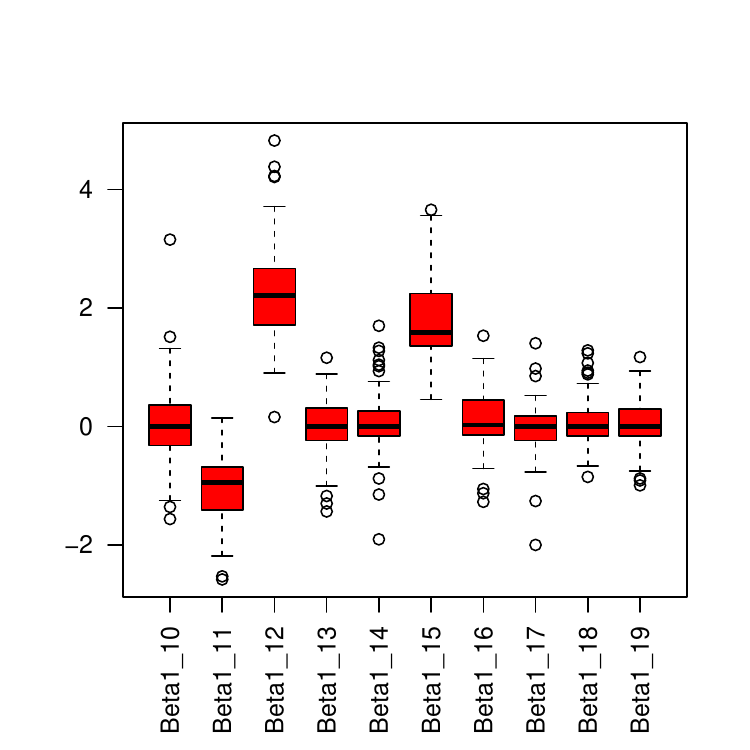}&
\includegraphics[width = 5 cm, height = 5 cm]{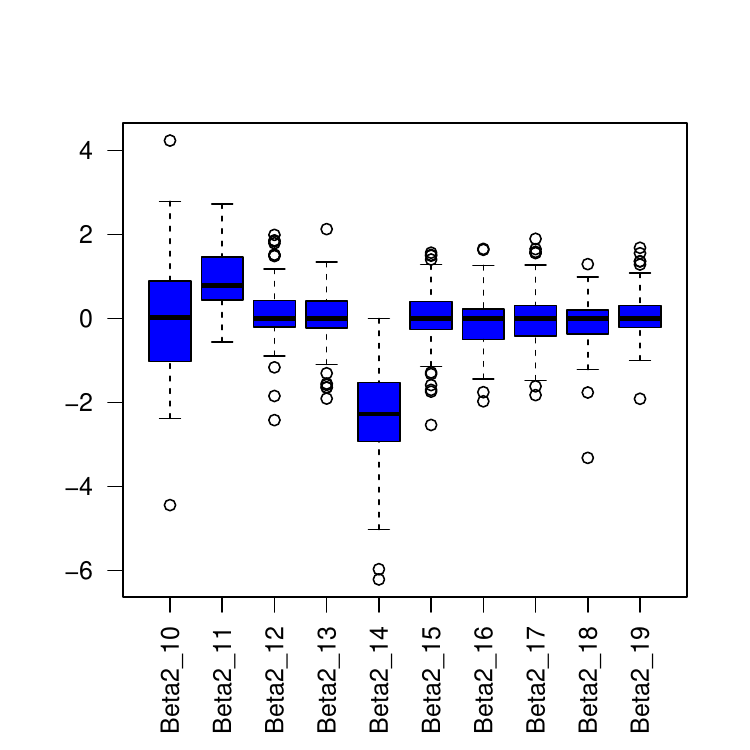}&
\includegraphics[width = 5 cm, height = 5 cm]{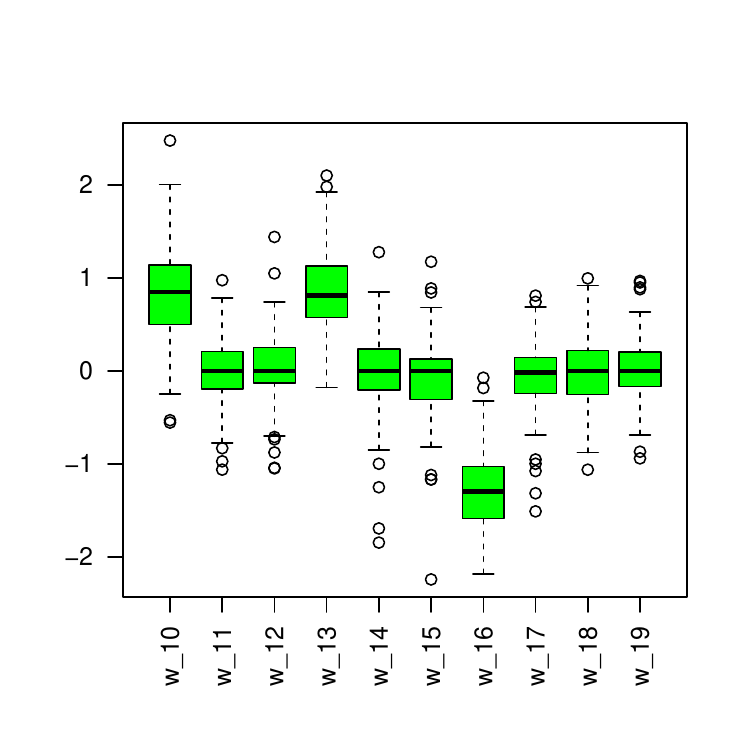}\\
MMoE+$\ell_2$-Exp.1 & MMoE+$\ell_2$-Exp.2 & MMoE+$\ell_2$-Gate\\
\includegraphics[width = 5 cm, height = 5 cm]{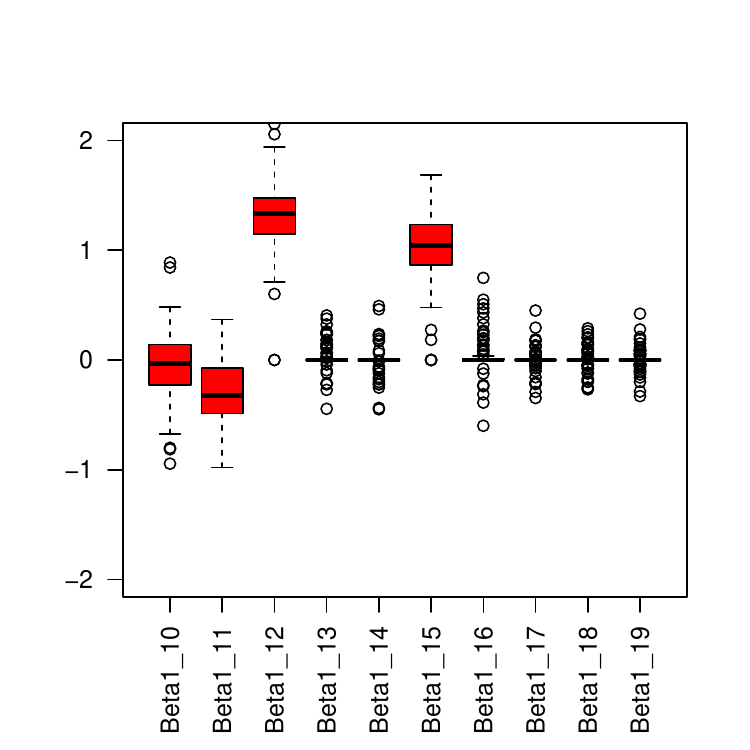}&
\includegraphics[width = 5 cm, height = 5 cm]{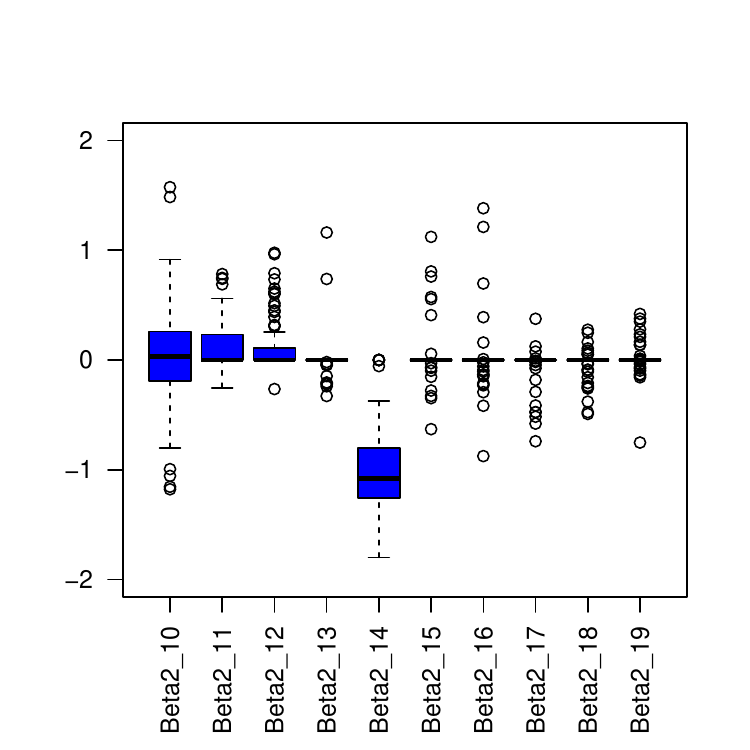}&
\includegraphics[width = 5 cm, height = 5 cm]{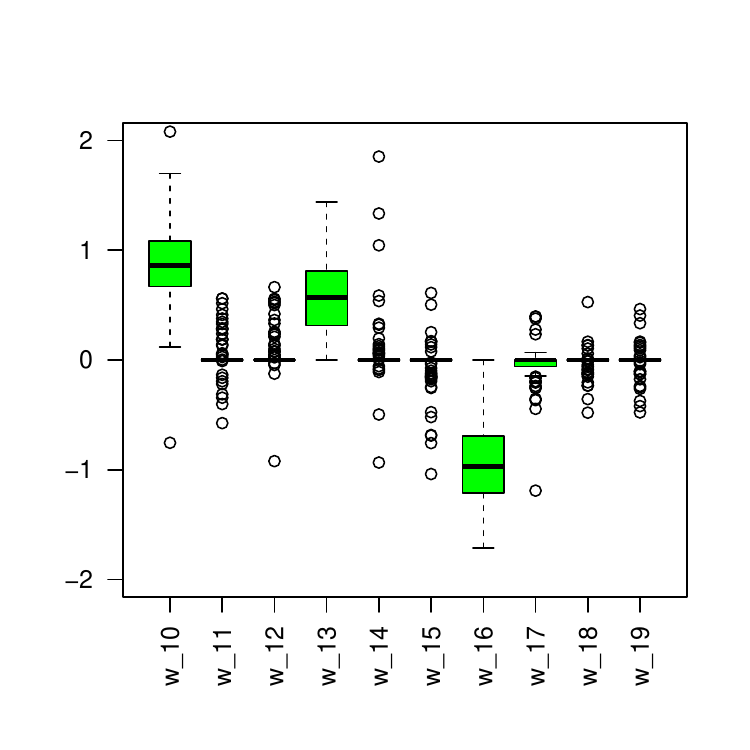}\\
MMoE-Lasso-Exp.1 & MMoE-Lasso-Exp.2 & MMoE-Lasso-Gate
\end{tabular}
\caption{Boxplots of $\ell_2$ regularized MoE model (MMoE+$\ell_2$) and the proposed Lasso penalized MoE model (MMoE-Lasso) for multinomial case.\label{Fig:Multinomial}}
\end{figure*}

The boxplots of all estimated parameters are shown in Figure \ref{Fig:Gaussian}, Figure \ref{Fig:Poisson} and Figure \ref{Fig:Multinomial}. Boxplots are not provided for the standard multinomial
model, since estimation of its parameters is unstable in this setting. The
non-penalized models GMoE, PMoE and MMoE+$\ell_2$ cannot be regarded as
model-selection procedures. In contrast, the Lasso yields sparse estimates in
both the experts and the gating network, and performs well in detecting nonzero coefficients, although this task is more difficult in the multinomial case.

\vk

The means and standard deviations are reported in Tables~\ref{Tab:EstimationGaussian}, \ref{Tab:EstimationPoisson} and \ref{Tab:EstimationMultinomial}. For the Gaussian and Poisson outputs, the standard (unpenalized) MoE gives better estimates of the nonzero coefficients than the $\ell_1$-regularized models; this is expected, since the penalty introduces bias into the estimated parameters. For the zero coefficients, however, GMoE-Lasso and PMoE-Lasso outperform GMoE and PMoE in terms of average mean squared error.\vk 

For the multinomial model, adding two small quadratic penalties removes the
instability encountered when estimating the MLE. For the nonzero coefficients, MMoE+$\ell_2$ and MMoE-Lasso show mixed performance, each outperforming the other on some coefficients. For the zero coefficients, as in the Gaussian and Poisson cases, the Lasso is effective in identifying zeros and yields smaller MSEs than MMoE+$\ell_2$.

\begin{table*}[!h]
\footnotesize
\centering
\begin{tabular}{|c|c|c|c||c|c|}
\hline
Comp. & True & \multicolumn{2}{c||}{Mean} & \multicolumn{2}{c|}{Mean squared error}\\
\cline{3-6}
& value & GMoE & GMoE-Lasso & GMoE & GMoE-Lasso \\
\hline
& 0 & $-0.005_{(\scriptstyle 0.098)}$ & $0.052_{(\scriptstyle 0.088)}$ & $0.010_{(\scriptstyle 0.001)}$ & $0.010_{(\scriptstyle 0.002)}$ \\ 
  & 0 & $0.008_{(\scriptstyle 0.104)}$ & $0.010_{(\scriptstyle 0.037)}$ & $0.011_{(\scriptstyle 0.002)}$ & $0.001_{(\scriptstyle 0.001)}$ \\ 
  & 1.5 & $1.509_{(\scriptstyle 0.091)}$ & $1.397_{(\scriptstyle 0.087)}$ & $0.008_{(\scriptstyle 0.001)}$ & $0.018_{(\scriptstyle 0.002)}$ \\ 
  & 0 & $0.008_{(\scriptstyle 0.096)}$ & $0.020_{(\scriptstyle 0.045)}$ & $0.009_{(\scriptstyle 0.001)}$ & $0.002_{(\scriptstyle 0.001)}$ \\ 
  Exp.1 & 0 & $-0.002_{(\scriptstyle 0.108)}$ & $0.013_{(\scriptstyle 0.033)}$ & $0.012_{(\scriptstyle 0.001)}$ & $0.001_{(\scriptstyle 0.000)}$ \\ 
  & 0 & $-0.001_{(\scriptstyle 0.100)}$ & $0.006_{(\scriptstyle 0.021)}$ & $0.010_{(\scriptstyle 0.001)}$ & $0.000_{(\scriptstyle 0.000)}$ \\ 
  & 1 & $0.993_{(\scriptstyle 0.105)}$ & $0.864_{(\scriptstyle 0.095)}$ & $0.011_{(\scriptstyle 0.001)}$ & $0.027_{(\scriptstyle 0.003)}$ \\ 
  & 0 & $0.010_{(\scriptstyle 0.111)}$ & $0.009_{(\scriptstyle 0.034)}$ & $0.012_{(\scriptstyle 0.002)}$ & $0.001_{(\scriptstyle 0.000)}$ \\ 
  & 0 & $-0.012_{(\scriptstyle 0.109)}$ & $-0.007_{(\scriptstyle 0.023)}$ & $0.012_{(\scriptstyle 0.002)}$ & $0.001_{(\scriptstyle 0.000)}$ \\ 
  & -2 & $-2.005_{(\scriptstyle 0.106)}$ & $-1.910_{(\scriptstyle 0.093)}$ & $0.011_{(\scriptstyle 0.002)}$ & $0.017_{(\scriptstyle 0.002)}$ \\ \hline
  & 0 & $-0.019_{(\scriptstyle 0.196)}$ & $-0.324_{(\scriptstyle 0.189)}$ & $0.038_{(\scriptstyle 0.005)}$ & $0.140_{(\scriptstyle 0.014)}$ \\ 
  & 1 & $0.990_{(\scriptstyle 0.157)}$ & $0.432_{(\scriptstyle 0.205)}$ & $0.025_{(\scriptstyle 0.004)}$ & $0.364_{(\scriptstyle 0.025)}$ \\ 
  & -1.5 & $-1.510_{(\scriptstyle 0.150)}$ & $-1.081_{(\scriptstyle 0.180)}$ & $0.023_{(\scriptstyle 0.003)}$ & $0.207_{(\scriptstyle 0.017)}$ \\ 
  & 0 & $0.028_{(\scriptstyle 0.151)}$ & $-0.001_{(\scriptstyle 0.022)}$ & $0.023_{(\scriptstyle 0.003)}$ & $0.000_{(\scriptstyle 0.000)}$ \\ 
  Exp.2  & 0 & $-0.016_{(\scriptstyle 0.161)}$ & $0.009_{(\scriptstyle 0.036)}$ & $0.026_{(\scriptstyle 0.004)}$ & $0.001_{(\scriptstyle 0.001)}$ \\ 
  & 2 & $1.995_{(\scriptstyle 0.162)}$ & $1.778_{(\scriptstyle 0.143)}$ & $0.026_{(\scriptstyle 0.006)}$ & $0.070_{(\scriptstyle 0.009)}$ \\ 
  & 0 & $0.012_{(\scriptstyle 0.164)}$ & $0.011_{(\scriptstyle 0.053)}$ & $0.027_{(\scriptstyle 0.004)}$ & $0.003_{(\scriptstyle 0.002)}$ \\ 
  & 0 & $0.018_{(\scriptstyle 0.158)}$ & $0.003_{(\scriptstyle 0.023)}$ & $0.025_{(\scriptstyle 0.003)}$ & $0.001_{(\scriptstyle 0.000)}$ \\ 
  & 0 & $-0.003_{(\scriptstyle 0.145)}$ & $-0.000_{(\scriptstyle 0.024)}$ & $0.021_{(\scriptstyle 0.003)}$ & $0.001_{(\scriptstyle 0.000)}$ \\ 
  & 0 & $-0.003_{(\scriptstyle 0.133)}$ & $0.006_{(\scriptstyle 0.030)}$ & $0.018_{(\scriptstyle 0.002)}$ & $0.001_{(\scriptstyle 0.001)}$ \\ \hline
  & 1 & $1.112_{(\scriptstyle 0.286)}$ & $0.690_{(\scriptstyle 0.144)}$ & $0.094_{(\scriptstyle 0.018)}$ & $0.116_{(\scriptstyle 0.009)}$ \\ 
  & 2 & $2.247_{(\scriptstyle 0.466)}$ & $1.039_{(\scriptstyle 0.159)}$ & $0.276_{(\scriptstyle 0.049)}$ & $0.949_{(\scriptstyle 0.030)}$ \\ 
  & 0 & $-0.004_{(\scriptstyle 0.303)}$ & $0.005_{(\scriptstyle 0.028)}$ & $0.091_{(\scriptstyle 0.013)}$ & $0.001_{(\scriptstyle 0.000)}$ \\ 
  & 0 & $-0.012_{(\scriptstyle 0.376)}$ & $-0.001_{(\scriptstyle 0.009)}$ & $0.14_{(\scriptstyle 0.019)}$ & $0.000_{(\scriptstyle 0.000)}$ \\ 
  Gate & -1 & $-1.128_{(\scriptstyle 0.352)}$ & $-0.218_{(\scriptstyle 0.143)}$ & $0.139_{(\scriptstyle 0.023)}$ & $0.631_{(\scriptstyle 0.022)}$ \\ 
  & 0 & $0.011_{(\scriptstyle 0.323)}$ & $-0.000_{(\scriptstyle 0.009)}$ & $0.104_{(\scriptstyle 0.015)}$ & $0.000_{(\scriptstyle 0.000)}$ \\ 
  & 1 & $1.149_{(\scriptstyle 0.416)}$ & $0.265_{(\scriptstyle 0.166)}$ & $0.193_{(\scriptstyle 0.031)}$ & $0.568_{(\scriptstyle 0.023)}$ \\ 
  & 0 & $-0.005_{(\scriptstyle 0.330)}$ & $0.015_{(\scriptstyle 0.052)}$ & $0.108_{(\scriptstyle 0.012)}$ & $0.003_{(\scriptstyle 0.001)}$ \\ 
  & 0 & $-0.037_{(\scriptstyle 0.288)}$ & $0.002_{(\scriptstyle 0.014)}$ & $0.084_{(\scriptstyle 0.010)}$ & $0.000_{(\scriptstyle 0.000)}$ \\ 
  & 0 & $0.029_{(\scriptstyle 0.251)}$ & $0.001_{(\scriptstyle 0.005)}$ & $0.063_{(\scriptstyle 0.010)}$ & $0.000_{(\scriptstyle 0.000)}$ \\ \hline
 $\sigma$ & 1 & $0.927_{(\scriptstyle 0.091)}$ & $1.068_{(\scriptstyle 0.135)}$ & $0.014_{(\scriptstyle 0.001)}$ & $0.023_{(\scriptstyle 0.006)}$ \\ 
\hline
\end{tabular}
\caption{Estimated parameter vector obtained by GMoE and our GMoE-Lasso for Gaussian outputs. \label{Tab:EstimationGaussian}}
\end{table*}
\begin{table*}[!h]
\footnotesize
\centering
\begin{tabular}{|c|c|c|c||c|c|}
\hline
Comp. & True & \multicolumn{2}{c||}{Mean} & \multicolumn{2}{c|}{Mean squared error}\\
\cline{3-6}
& value & PMoE & PMoE-Lasso & PMoE & PMoE-Lasso \\
\hline
& 0 & $-0.009_{(\scriptstyle 0.108)}$ & $0.271_{(\scriptstyle 0.101)}$ & $0.012_{(\scriptstyle 0.002)}$ & $0.083_{(\scriptstyle 0.006)}$ \\ 
&  1 & $0.993_{(\scriptstyle 0.087)}$ & $0.858_{(\scriptstyle 0.092)}$ & $0.008_{(\scriptstyle 0.001)}$ & $0.029_{(\scriptstyle 0.003)}$ \\ 
&  0 & $-0.008_{(\scriptstyle 0.091)}$ & $-0.004_{(\scriptstyle 0.016)}$ & $0.008_{(\scriptstyle 0.001)}$ & $0.000_{(\scriptstyle 0.000)}$ \\ 
&  -2 & $-1.986_{(\scriptstyle 0.100)}$ & $-1.783_{(\scriptstyle 0.111)}$ & $0.010_{(\scriptstyle 0.002)}$ & $0.059_{(\scriptstyle 0.005)}$ \\ 
Exp.1 &  0 & $-0.015_{(\scriptstyle 0.080)}$ & $-0.000_{(\scriptstyle 0.018)}$ & $0.007_{(\scriptstyle 0.001)}$ & $0.000_{(\scriptstyle 0.000)}$ \\ 
&  1.5 & $1.495_{(\scriptstyle 0.106)}$ & $1.298_{(\scriptstyle 0.096)}$ & $0.011_{(\scriptstyle 0.002)}$ & $0.050_{(\scriptstyle 0.004)}$ \\ 
&  0 & $-0.004_{(\scriptstyle 0.087)}$ & $0.003_{(\scriptstyle 0.020)}$ & $0.008_{(\scriptstyle 0.001)}$ & $0.000_{(\scriptstyle 0.000)}$ \\ 
&  0 & $0.015_{(\scriptstyle 0.103)}$ & $-0.008_{(\scriptstyle 0.027)}$ & $0.011_{(\scriptstyle 0.002)}$ & $0.001_{(\scriptstyle 0.000)}$ \\ 
&  0 & $0.003_{(\scriptstyle 0.072)}$ & $-0.001_{(\scriptstyle 0.023)}$ & $0.005_{(\scriptstyle 0.001)}$ & $0.001_{(\scriptstyle 0.000)}$ \\ 
&  0 & $0.001_{(\scriptstyle 0.075)}$ & $-0.004_{(\scriptstyle 0.022)}$ & $0.006_{(\scriptstyle 0.001)}$ & $0.000_{(\scriptstyle 0.000)}$ \\ \hline
&  0 & $-0.052_{(\scriptstyle 0.466)}$ & $0.269_{(\scriptstyle 0.161)}$ & $0.218_{(\scriptstyle 0.161)}$ & $0.098_{(\scriptstyle 0.010)}$ \\ 
&  0 & $0.033_{(\scriptstyle 0.180)}$ & $0.020_{(\scriptstyle 0.080)}$ & $0.033_{(\scriptstyle 0.011)}$ & $0.007_{(\scriptstyle 0.005)}$ \\ 
&  2 & $1.967_{(\scriptstyle 0.462)}$ & $1.772_{(\scriptstyle 0.171)}$ & $0.212_{(\scriptstyle 0.167)}$ & $0.081_{(\scriptstyle 0.012)}$ \\ 
&  0 & $-0.048_{(\scriptstyle 0.232)}$ & $-0.005_{(\scriptstyle 0.030)}$ & $0.056_{(\scriptstyle 0.029)}$ & $0.001_{(\scriptstyle 0.001)}$ \\ 
Exp.2 &  -1 & $-0.963_{(\scriptstyle 0.302)}$ & $-0.793_{(\scriptstyle 0.164)}$ & $0.092_{(\scriptstyle 0.043)}$ & $0.070_{(\scriptstyle 0.012)}$ \\ 
&  0 & $0.048_{(\scriptstyle 0.237)}$ & $-0.006_{(\scriptstyle 0.021)}$ & $0.058_{(\scriptstyle 0.023)}$ & $0.000_{(\scriptstyle 0.000)}$ \\ 
&  0 & $-0.057_{(\scriptstyle 0.301)}$ & $-0.008_{(\scriptstyle 0.071)}$ & $0.093_{(\scriptstyle 0.050)}$ & $0.005_{(\scriptstyle 0.004)}$ \\ 
&  0 & $-0.084_{(\scriptstyle 0.330)}$ & $0.003_{(\scriptstyle 0.037)}$ & $0.115_{(\scriptstyle 0.079)}$ & $0.001_{(\scriptstyle 0.001)}$ \\ 
&  0 & $0.063_{(\scriptstyle 0.262)}$ & $0.001_{(\scriptstyle 0.016)}$ & $0.072_{(\scriptstyle 0.036)}$ & $0.000_{(\scriptstyle 0.000)}$ \\ 
&  0 & $-0.053_{(\scriptstyle 0.271)}$ & $0.005_{(\scriptstyle 0.054)}$ & $0.076_{(\scriptstyle 0.048)}$ & $0.003_{(\scriptstyle 0.003)}$ \\ \hline
&  1 & $1.128_{(\scriptstyle 0.434)}$ & $0.601_{(\scriptstyle 0.179)}$ & $0.203_{(\scriptstyle 0.070)}$ & $0.191_{(\scriptstyle 0.014)}$ \\ 
&  0 & $0.034_{(\scriptstyle 0.319)}$ & $0.003_{(\scriptstyle 0.014)}$ & $0.102_{(\scriptstyle 0.019)}$ & $0.000_{(\scriptstyle 0.000)}$ \\ 
&  0 & $-0.026_{(\scriptstyle 0.353)}$ & $0.080_{(\scriptstyle 0.114)}$ & $0.124_{(\scriptstyle 0.022)}$ & $0.019_{(\scriptstyle 0.004)}$ \\ 
&  1 & $1.167_{(\scriptstyle 0.421)}$ & $0.464_{(\scriptstyle 0.163)}$ & $0.203_{(\scriptstyle 0.039)}$ & $0.314_{(\scriptstyle 0.018)}$ \\ 
Gate &  0 & $-0.062_{(\scriptstyle 0.334)}$ & $-0.002_{(\scriptstyle 0.013)}$ & $0.114_{(\scriptstyle 0.025)}$ & $0.000_{(\scriptstyle 0.000)}$ \\ 
&  -1.5 & $-1.637_{(\scriptstyle 0.433)}$ & $-0.793_{(\scriptstyle 0.153)}$ & $0.204_{(\scriptstyle 0.028)}$ & $0.523_{(\scriptstyle 0.023)}$ \\ 
&  0 & $-0.008_{(\scriptstyle 0.315)}$ & $-0.025_{(\scriptstyle 0.072)}$ & $0.098_{(\scriptstyle 0.014)}$ & $0.006_{(\scriptstyle 0.003)}$ \\ 
&  -1 & $-1.138_{(\scriptstyle 0.404)}$ & $-0.424_{(\scriptstyle 0.170)}$ & $0.181_{(\scriptstyle 0.024)}$ & $0.361_{(\scriptstyle 0.020)}$ \\ 
&  0 & $0.003_{(\scriptstyle 0.325)}$ & $-0.011_{(\scriptstyle 0.038)}$ & $0.105_{(\scriptstyle 0.017)}$ & $0.002_{(\scriptstyle 0.001)}$ \\ 
&  0 & $0.035_{(\scriptstyle 0.300)}$ & $-0.002_{(\scriptstyle 0.015)}$ & $0.091_{(\scriptstyle 0.016)}$ & $0.000_{(\scriptstyle 0.000)}$ \\ 
\hline
\end{tabular}
\caption{Estimated parameter vector obtained by PMoE and our PMoE-Lasso for Poisson outputs. \label{Tab:EstimationPoisson}}
\end{table*}
\begin{table*}[!h]
\footnotesize
\centering
\begin{tabular}{|c|c|c|c||c|c|}
\hline
Comp. & True & \multicolumn{2}{c||}{Mean} & \multicolumn{2}{c|}{Mean squared error}\\
\cline{3-6}
& value & MMoE+$\ell_2$ & MMoE-Lasso & MMoE+$\ell_2$ & MMoE-Lasso \\
\hline
& 0 & $0.033_{(\scriptstyle 0.721)}$ & $-0.265_{(\scriptstyle 1.637)}$ & $0.516_{(\scriptstyle 0.111)}$ & $2.722_{(\scriptstyle 1.863)}$ \\ 
&  -1 & $-1.037_{(\scriptstyle 0.551)}$ & $-0.310_{(\scriptstyle 0.255)}$ & $0.302_{(\scriptstyle 0.044)}$ & $0.540_{(\scriptstyle 0.035)}$ \\ 
&  2 & $2.283_{(\scriptstyle 0.820)}$ & $1.300_{(\scriptstyle 0.324)}$ & $0.746_{(\scriptstyle 0.131)}$ & $0.594_{(\scriptstyle 0.062)}$ \\ 
&  0 & $-0.015_{(\scriptstyle 0.463)}$ & $0.024_{(\scriptstyle 0.116)}$ & $0.212_{(\scriptstyle 0.035)}$ & $0.014_{(\scriptstyle 0.003)}$ \\ 
Exp.1 &  0 & $0.063_{(\scriptstyle 0.486)}$ & $-0.002_{(\scriptstyle 0.119)}$ & $0.238_{(\scriptstyle 0.055)}$ & $0.014_{(\scriptstyle 0.004)}$ \\ 
&  1.5 & $1.776_{(\scriptstyle 0.625)}$ & $1.023_{(\scriptstyle 0.335)}$ & $0.463_{(\scriptstyle 0.077)}$ & $0.339_{(\scriptstyle 0.045)}$ \\ 
&  0 & $0.128_{(\scriptstyle 0.489)}$ & $0.035_{(\scriptstyle 0.167)}$ & $0.253_{(\scriptstyle 0.040)}$ & $0.029_{(\scriptstyle 0.008)}$ \\ 
&  0 & $-0.039_{(\scriptstyle 0.409)}$ & $0.001_{(\scriptstyle 0.095)}$ & $0.167_{(\scriptstyle 0.048)}$ & $0.009_{(\scriptstyle 0.003)}$ \\ 
&  0 & $0.047_{(\scriptstyle 0.383)}$ & $0.005_{(\scriptstyle 0.093)}$ & $0.147_{(\scriptstyle 0.029)}$ & $0.009_{(\scriptstyle 0.002)}$ \\ 
&  0 & $0.037_{(\scriptstyle 0.404)}$ & $0.005_{(\scriptstyle 0.090)}$ & $0.163_{(\scriptstyle 0.025)}$ & $0.008_{(\scriptstyle 0.002)}$ \\ \hline
&  0 & $0.054_{(\scriptstyle 1.328)}$ & $0.055_{(\scriptstyle 0.920)}$ & $1.748_{(\scriptstyle 0.289)}$ & $0.841_{(\scriptstyle 0.486)}$ \\ 
&  1 & $0.955_{(\scriptstyle 0.757)}$ & $0.129_{(\scriptstyle 0.207)}$ & $4.391_{(\scriptstyle 0.321)}$ & $1.318_{(\scriptstyle 0.053)}$ \\ 
&  0 & $0.146_{(\scriptstyle 0.719)}$ & $0.106_{(\scriptstyle 0.229)}$ & $3.950_{(\scriptstyle 0.278)}$ & $3.639_{(\scriptstyle 0.076)}$ \\ 
&  0 & $-0.014_{(\scriptstyle 0.654)}$ & $0.004_{(\scriptstyle 0.150)}$ & $0.424_{(\scriptstyle 0.080)}$ & $0.022_{(\scriptstyle 0.015)}$ \\ 
Exp.2 &  -2 & $-2.290_{(\scriptstyle 1.086)}$ & $-1.015_{(\scriptstyle 0.421)}$ & $6.411_{(\scriptstyle 0.625)}$ & $1.205_{(\scriptstyle 0.083)}$ \\ 
&  0 & $-0.009_{(\scriptstyle 0.710)}$ & $0.023_{(\scriptstyle 0.201)}$ & $2.775_{(\scriptstyle 0.255)}$ & $2.222_{(\scriptstyle 0.05)}$ \\ 
&  0 & $-0.107_{(\scriptstyle 0.659)}$ & $0.009_{(\scriptstyle 0.231)}$ & $0.441_{(\scriptstyle 0.072)}$ & $0.053_{(\scriptstyle 0.025)}$ \\ 
&  0 & $-0.020_{(\scriptstyle 0.703)}$ & $-0.032_{(\scriptstyle 0.141)}$ & $0.490_{(\scriptstyle 0.078)}$ & $0.021_{(\scriptstyle 0.008)}$ \\ 
&  0 & $-0.097_{(\scriptstyle 0.578)}$ & $-0.018_{(\scriptstyle 0.103)}$ & $0.340_{(\scriptstyle 0.116)}$ & $0.011_{(\scriptstyle 0.004)}$ \\ 
&  0 & $0.056_{(\scriptstyle 0.561)}$ & $0.009_{(\scriptstyle 0.116)}$ & $0.315_{(\scriptstyle 0.061)}$ & $0.013_{(\scriptstyle 0.006)}$ \\ \hline
&  1 & $0.851_{(\scriptstyle 0.517)}$ & $0.791_{(\scriptstyle 0.770)}$ & $0.287_{(\scriptstyle 0.048)}$ & $0.631_{(\scriptstyle 0.486)}$ \\ 
&  0 & $-0.010_{(\scriptstyle 0.338)}$ & $0.033_{(\scriptstyle 0.169)}$ & $0.113_{(\scriptstyle 0.021)}$ & $0.029_{(\scriptstyle 0.007)}$ \\ 
&  0 & $0.016_{(\scriptstyle 0.400)}$ & $0.053_{(\scriptstyle 0.180)}$ & $0.159_{(\scriptstyle 0.030)}$ & $0.035_{(\scriptstyle 0.011)}$ \\ 
&  1 & $0.887_{(\scriptstyle 0.505)}$ & $0.556_{(\scriptstyle 0.353)}$ & $0.266_{(\scriptstyle 0.032)}$ & $0.320_{(\scriptstyle 0.033)}$ \\ 
Gate &  0 & $-0.011_{(\scriptstyle 0.479)}$ & $0.056_{(\scriptstyle 0.288)}$ & $0.228_{(\scriptstyle 0.051)}$ & $0.085_{(\scriptstyle 0.041)}$ \\ 
&  0 & $-0.080_{(\scriptstyle 0.463)}$ & $-0.044_{(\scriptstyle 0.204)}$ & $0.219_{(\scriptstyle 0.057)}$ & $0.043_{(\scriptstyle 0.015)}$ \\ 
&  -1.5 & $-1.289_{(\scriptstyle 0.438)}$ & $-0.929_{(\scriptstyle 0.409)}$ & $0.234_{(\scriptstyle 0.035)}$ & $0.492_{(\scriptstyle 0.060)}$ \\ 
&  0 & $-0.071_{(\scriptstyle 0.385)}$ & $-0.038_{(\scriptstyle 0.169)}$ & $0.152_{(\scriptstyle 0.034)}$ & $0.030_{(\scriptstyle 0.015)}$ \\ 
&  0 & $-0.001_{(\scriptstyle 0.361)}$ & $-0.013_{(\scriptstyle 0.098)}$ & $0.129_{(\scriptstyle 0.021)}$ & $0.010_{(\scriptstyle 0.004)}$ \\ 
&  0 & $0.018_{(\scriptstyle 0.344)}$ & $-0.008_{(\scriptstyle 0.120)}$ & $0.118_{(\scriptstyle 0.021)}$ & $0.014_{(\scriptstyle 0.004)}$ \\ 
\hline
\end{tabular}
\caption{Estimated parameter vector obtained by MMoE+$\ell_2$ and the proposed MMoE-Lasso models for multinomial outputs. \label{Tab:EstimationMultinomial}}
\end{table*}
\subsubsection{Clustering}

The clustering accuracy of all the models is computed for each data set; the
results, in terms of the correct classification rate and the Adjusted Rand Index (ARI), are reported in Table~\ref{Cluster}. Overall, the differences between the penalized and non-penalized models are small---of the same order as the Monte Carlo variability, so that the Lasso attains clustering accuracy comparable to the unpenalized models while additionally delivering sparsity. For the Gaussian case, GMoE and GMoE-Lasso perform similarly, with GMoE slightly ahead (correct classification rate $91.0\%$ versus $89.9\%$; ARI $0.670$ versus $0.635$). For the Poisson case, PMoE-Lasso and PMoE are essentially indistinguishable, the difference being below $1\%$. For the multinomial case, MMoE-Lasso is marginally better than MMoE+$\ell_2$ in this simulation, although the large standard deviations indicate that this difference is not significant.

\begin{table}[!h]
\centering
{\small\begin{tabular}{|c|c|c|}
\hline
Model & C.rate & ARI\\
\hline
\hline
GMoE & $91.00\%_{(\scriptstyle 1.78\%)}$ & $0.670_{(\scriptstyle 0.059)}$ \\ 
GMoE-Lasso &  $89.92\%_{(\scriptstyle 2.06\%)}$ & $0.635_{(\scriptstyle 0.066)}$ \\ \hline
PMoE &  $88.64\%_{(\scriptstyle 2.58\%)}$ & $0.594_{(\scriptstyle 0.078)}$ \\ 
PMoE-Lasso &  $88.74\%_{(\scriptstyle 2.15\%)}$ & $0.597_{(\scriptstyle 0.066)}$ \\ \hline
MMoE+$\ell_2$ &  $78.54\%_{(\scriptstyle 5.37\%)}$ & $0.327_{(\scriptstyle 0.105)}$ \\ 
MMoE-Lasso &  $79.31\%_{(\scriptstyle 7.72\%)}$ & $0.350_{(\scriptstyle 0.116)}$ \\ 
\hline
\end{tabular}}
\caption{Average of the accuracy of clustering (correct classification rate and Adjusted Rand Index) results for the Gaussian and Poisson outputs with the non-penalized model (GMoE, PMoE) and the proposed Lasso regularized models (GMoE-Lasso, PMoE-Lasso). The results are also considered for the multinomial case with $\ell_2$ penalization (MMoE+$\ell_2$) and Lasso regularization (MMoE-Lasso).\label{Cluster}}
\end{table}\vk
It is clear that the regularized methods perform quite well in retrieving the actual sparse support; the sensitivity and specificity results are quite reasonable for the proposed models. Although the penalty function will cause bias to the parameters, as shown in the results of the MSE, the algorithm can  perform parameter density estimation  with an acceptable loss of information due to the bias induced by the regularization. 
In terms of clustering, the $\ell_1$ regularized models work as well as MoE models for the Gaussian and Poisson models. For multinomial model, the Lasso is successful in retrieving the actual parameters used for the model, while the non regularized method failed in this task. Hence, adding quadratic penalty functions in this case is necessary. A comparison between the MMoE-Lasso and the MMoE+$\ell_2$ model shows that Lasso provides better results not only in terms of sparsity but also in terms of clustering.
\subsection{Applications to real data sets}
In this part, five real data sets are analyzed as a further test of the proposed methodology. Two data sets are for the Gaussian model, two for the multinomial model and one for Poisson model. The obtain results are compared with other methods, which provided by \cite{Kha10} and \cite{Per14}. The comparison are based upon three different criteria: the average mean squared error (MSE) between observation values and the predicted values of the response variable, the sparsity of each result, and the correlation of these values. After the parameters are estimated and the data are clustered, the following value under the estimated model
\begin{equation*}
\hat{Y} = \arg\max p_k(y|\bsx; \hat{z}=k) = \arg\max p_k(y|\bsx;\hat{\bstheta}_k),
 \end{equation*}
is used as a predicted value for $Y$. 
\subsubsection{MoE model with Gaussian outputs}
The regularized MoE for Gaussian model are tested on two real data sets: the housing data and the residential building data described on the website UC Irvine Machine Learning Repository. This was done to provide a comparison with the experiment of \cite{Kha10} on housing data.\vk
\begin{figure*}[!h]
\centering
\begin{tabular}{ccc}
\includegraphics[width = 5cm, height = 5cm]{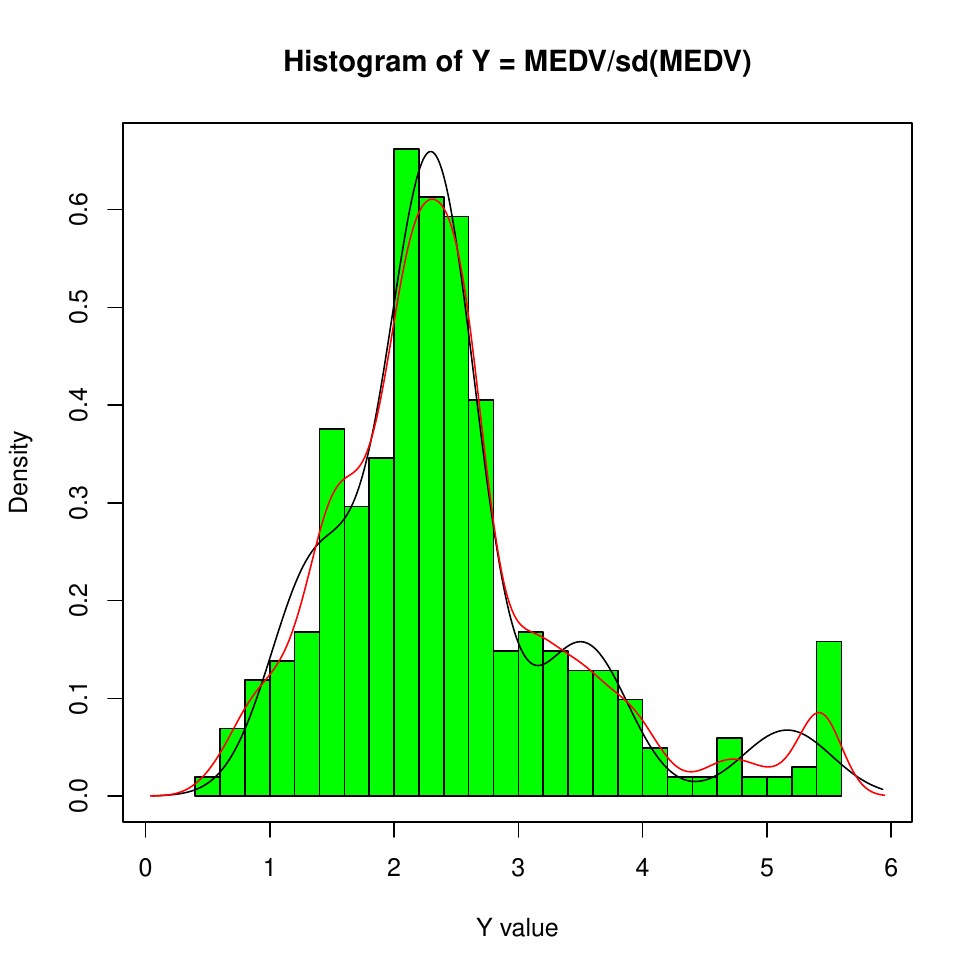}&
\includegraphics[width = 5cm, height = 5cm]{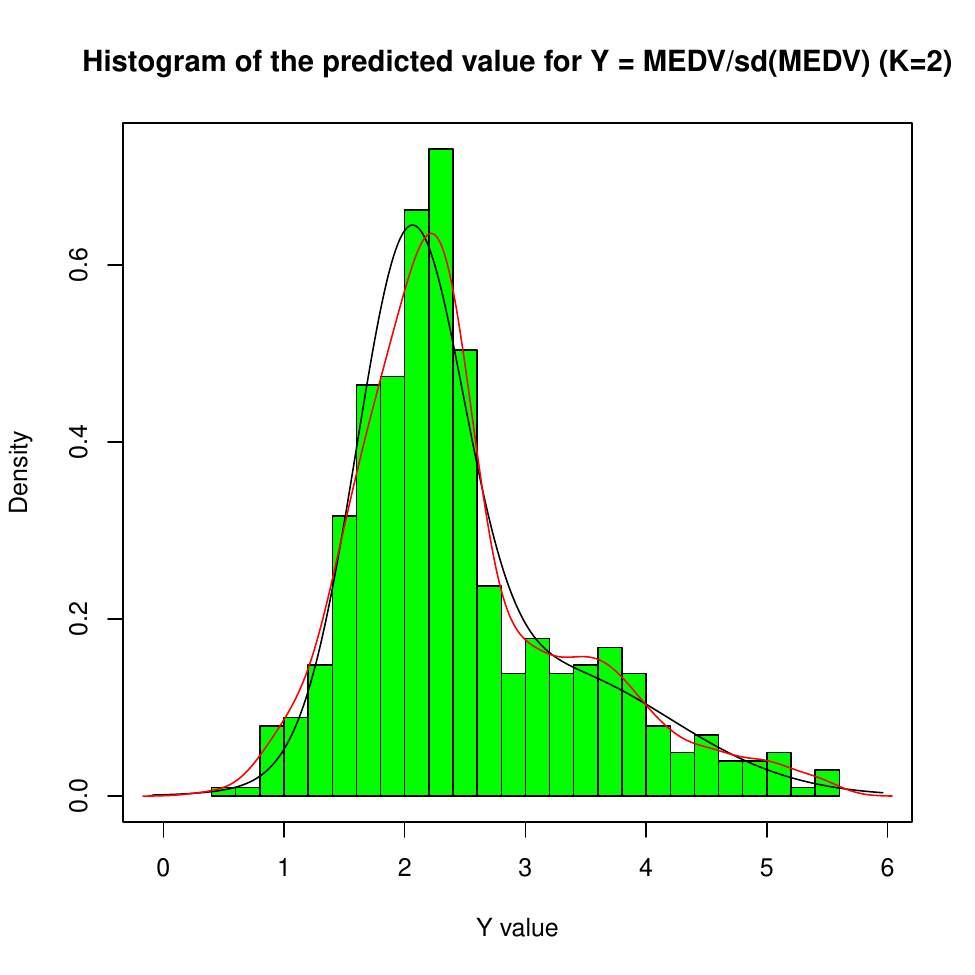}&
\includegraphics[width = 5cm, height = 5cm]{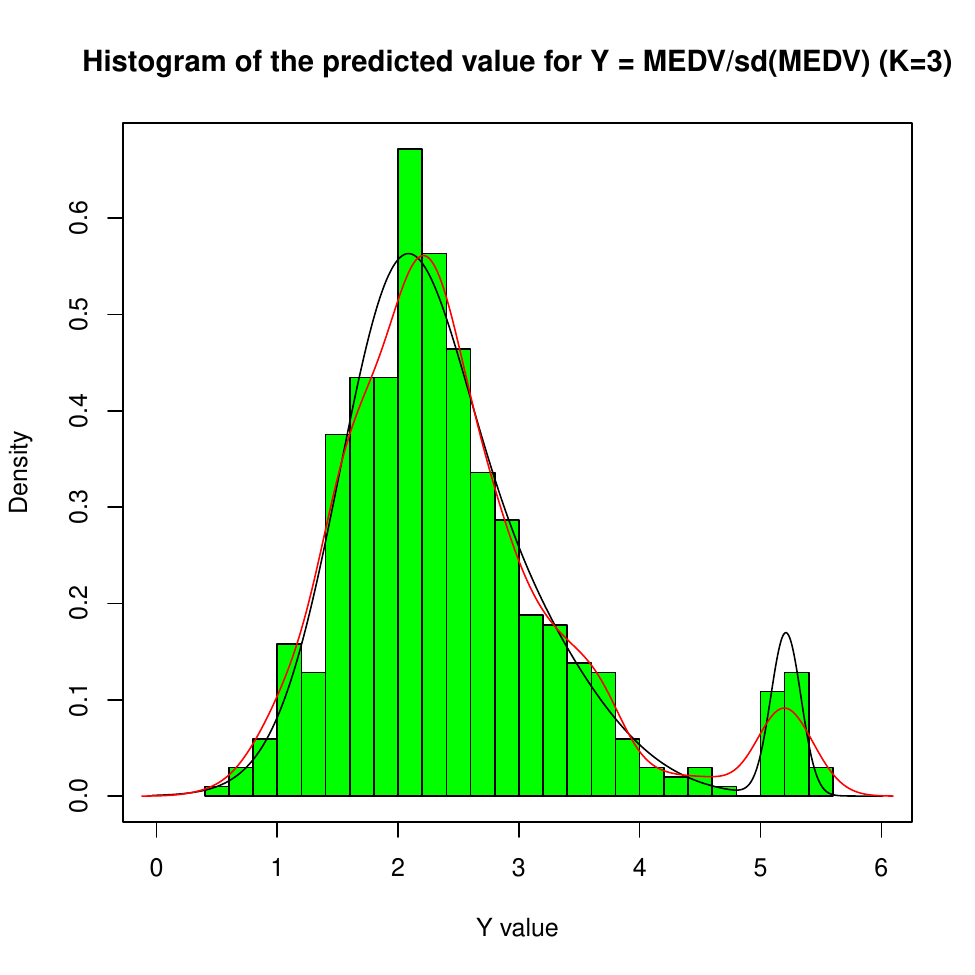}\\
 Histogram of $Y$ & Histogram of $\hat{Y} (K=2)$ & Histogram of $\hat{Y} (K=3)$ 
\end{tabular}
\vspace{-.4cm}
\caption{Histograms of $Y$ = MEDV/sd(MEDV) and its predicted value for housing data. The red lines represent the estimated densities using kernel density estimation and the black lines show the estimated densities using a GMM.\label{HousingDataHistogram}} 
\end{figure*}
The housing data set concerns houses' value in the suburbs of Boston. It has $506$ observations and $13$ features that may affect the houses' value.  The columns of $X$ were standardized to have the mean equal to $0$ and the variance equal to $1$. The response variable of interest is the median value of owner occupied homes in $\$1000's$, MEDV. Based on the histogram of $Y$ = MEDV/sd(MEDV), where sd(MEDV) is the standard deviation of MEDV, Khalili separated $Y$ into two groups of houses with $``$low$"$ and $``$high$"$ values. Hence, a MoE model is used to  fit the response
$$Y\sim\pi_1(\bsx;\bw)\mathcal{N}(y; \beta_{10}+ \bsx^\top\bsbeta_1,\sigma^2) + (1-\pi_1(\bsx;\bw))\mathcal{N}(y; \beta_{20}+ \bsx^\top\bsbeta_2,\sigma^2),$$ 
where $\displaystyle\pi_1(\bsx;\bw) = \frac{e^{w_{10}+\bsx^\top \bsw_1}}{1+e^{w_{10}+\bsx^\top \bsw_1}}$.
The estimated parameter of the MoE models obtained by Lasso and MLE are given in Table \ref{House}. These results are compared with Khalili's results. 
\begin{table}[!h]
\footnotesize
\begin{center}
\begin{tabular}{|c|c|c|c||c|c|c|}
\hline
Features & \multicolumn{3}{|c||}{GMoE-Lasso+$\ell_2$ (Khalili), $\sigma = 0.352$} & \multicolumn{3}{c|}{GMoE-Lasso, $\sigma = 0.353$}\\
\cline{2-7}
\ & Exp.1 & Exp.2 & Gate& Exp.1 & Exp.2 & Gate\\
\hline
$x_{0}$	&2.16	&2.84	&1.04	&2.18859	&2.82834	&1.00241\\
$x_{1}$	&-0.09	&-	&-	&-0.08818	&-	&-\\
$x_{2}$	&-	&0.07	&-	&-	&0.06312	&-\\
$x_{3}$	&-	&-	&0.67	&-	&-	&0.58559\\
$x_{4}$	&-	&0.05	&-	&0.04189	&0.05606	&-\\
$x_{5}$	&-	&-	&-	&-0.06550	&-	&-\\
$x_{6}$	&-	&0.60	&-0.27	&-	&0.58868	&-0.20882\\
$x_{7}$	&-	&-	&-	&-0.03640	&-	&-\\
$x_{8}$	&-	&-0.20	&-	&-	&-0.19447	&-\\
$x_{9}$	&-	&0.55	&-	&-	&0.54518	&-\\
$x_{10}$	&-	&-	&-	&-0.00329	&-	&-\\
$x_{11}$	&-	&-	&0.54	&-0.08641	&-0.06184	&0.39455\\
$x_{12}$	&0.05	&-	&-	&0.05058	&-	&-\\
$x_{13}$	&-0.29	&-0.49	&1.56	&-0.29022	&-0.50688	&1.36238\\
\hline
\end{tabular}
\caption{Fitted models for housing data with GMoE-Lasso+$\ell_2$ \citep{Kha10} and our proposed algorithm GMoE-Lasso.\label{House}}
\end{center}
\end{table}
In Table \ref{ResultHousing}, the results are provided in terms of average MSE and the correlation between the true observation value $Y$ and its prediction $\hat{Y}$. A few parameters in both methods have the same value. The MSE and the correlation from the proposed method are better than those in \cite{Kha10}. 
\begin{table}[!h]
    \centering
    \begin{tabular}{|c|c|c|}
    \hline
          &GMoE-Lasso+$\ell_2$ (Khalili) & GMoE-Lasso \\
         \hline
        $R^2$ & 0.8698 & 0.8832\\
        \hline
        MSE & $0.1371_{(.286)}$ & $0.1178_{(.282)}$\\
        \hline
    \end{tabular}
    \caption{Results for Housing data set with a two-component MoE model.}
    \label{ResultHousing}
\end{table}\vk
Considering the case $K=3$ as an extension. The estimated parameters, the average MSE, and the correlation between the true observation value $Y$ and its prediction $\hat{Y}$ for this case can be found in Table \ref{HouseK=3} and Table \ref{ResultHousingK=3}. It turns out that this model provides better results than those with $K=2$ in terms of prediction. The modified BIC increases from $-292.822 \, (K=2)$ to $-246.844\, (K=3)$.

The histograms of $\hat{Y}$ for $K=2$ and $K=3$ are also compared with the histogram of $Y$ (see Figure \ref{HousingDataHistogram}). It turns out that with $K=3$ the histogram of $\hat{Y}$ provides better result in approximating the histogram of the true value. 
\begin{table}[!h]
\footnotesize
\begin{center}
\begin{tabular}{|c|c|c|c||c|c|}
\hline
Features & \multicolumn{3}{c||}{Expert,\ $\sigma = 0.261$}&\multicolumn{2}{c|}{Gating network}\\
\cline{2-6}
& Exp.1 & Exp.2 & Exp.3 & Gate.1 & Gate.2\\
\hline
$x_{0}$	&2.14331	&5.01278	&2.50307	&-0.27941	&-2.96191\\
$x_{1}$	&-0.09202	&-	&-	&0.01695	&-\\
$x_{2}$	&-	&0.03392	&0.01033	&-	&-\\
$x_{3}$	&-	&-	&-0.03802	&-	&-\\
$x_{4}$	&0.05261	&0.01517	&0.00950	&-	&0.12079\\
$x_{5}$	&-0.12082	&-	&-	&-	&-\\
$x_{6}$	&-0.08837	&0.12770	&0.67982	&-	&0.97405\\
$x_{7}$	&-	&-	&-0.17057	&0.27293	&-\\
$x_{8}$	&-0.08727	&-	&-0.12630	&-	&-0.27807\\
$x_{9}$	&0.04286	&-	&0.11111	&-	&-\\
$x_{10}$	&-0.06967	&0.21112	&-0.13565	&0.42344	&-\\
$x_{11}$	&-0.08817	&-	&-0.11758	&0.01711	&-0.02419\\
$x_{12}$	&0.03348	&-	&-	&-0.22068	&-\\
$x_{13}$	&-0.34326	&-	&-	&1.01512	&-\\  
\hline
\end{tabular}
\caption{Fitted models for housing data set with GMoE-Lasso method ($K=3$).\label{HouseK=3}}
\end{center}
\end{table}
\begin{table}[!h]
    \centering
    \begin{tabular}{|c|c|c||c|c|c|}
    \hline
	Method &\multicolumn{2}{c||}{Criteria} & \multicolumn{3}{c|}{Number of observations}\\
   \cline{2-6}
          & $R^2$ & MSE & Class 1 & Class 2 & Class 3\\
         \hline
        GMoE-Lasso ($K=3$) & 0.9372 & $0.0629_{(.106)}$ & 195 & 28 & 283\\
        \hline
    \end{tabular}
    \caption{Results for Housing data set obtained by our GMoE-Lasso method ($K=3$).}
    \label{ResultHousingK=3}
\end{table}\vk
\begin{figure}[!h]
\centering
\begin{tabular}{cc}
\includegraphics[width = 6.5 cm, height = 6.5 cm]{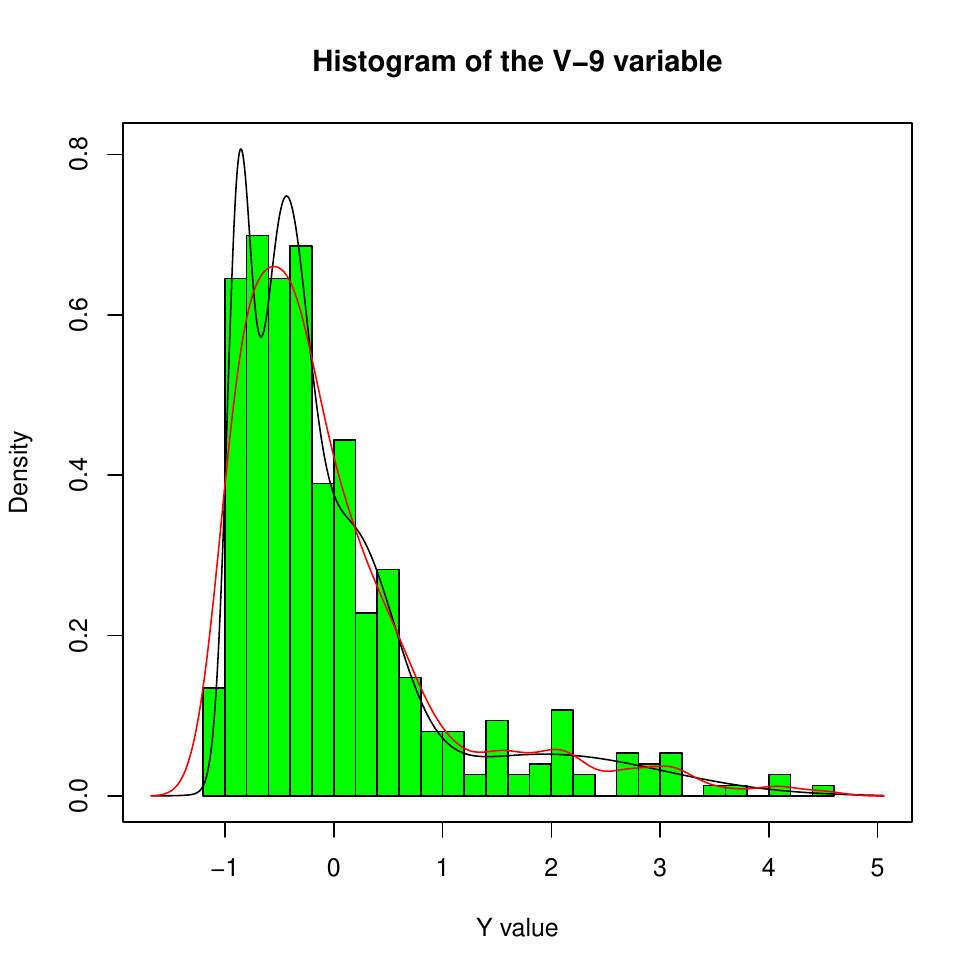}&
\includegraphics[width = 6.5 cm, height = 6.5 cm]{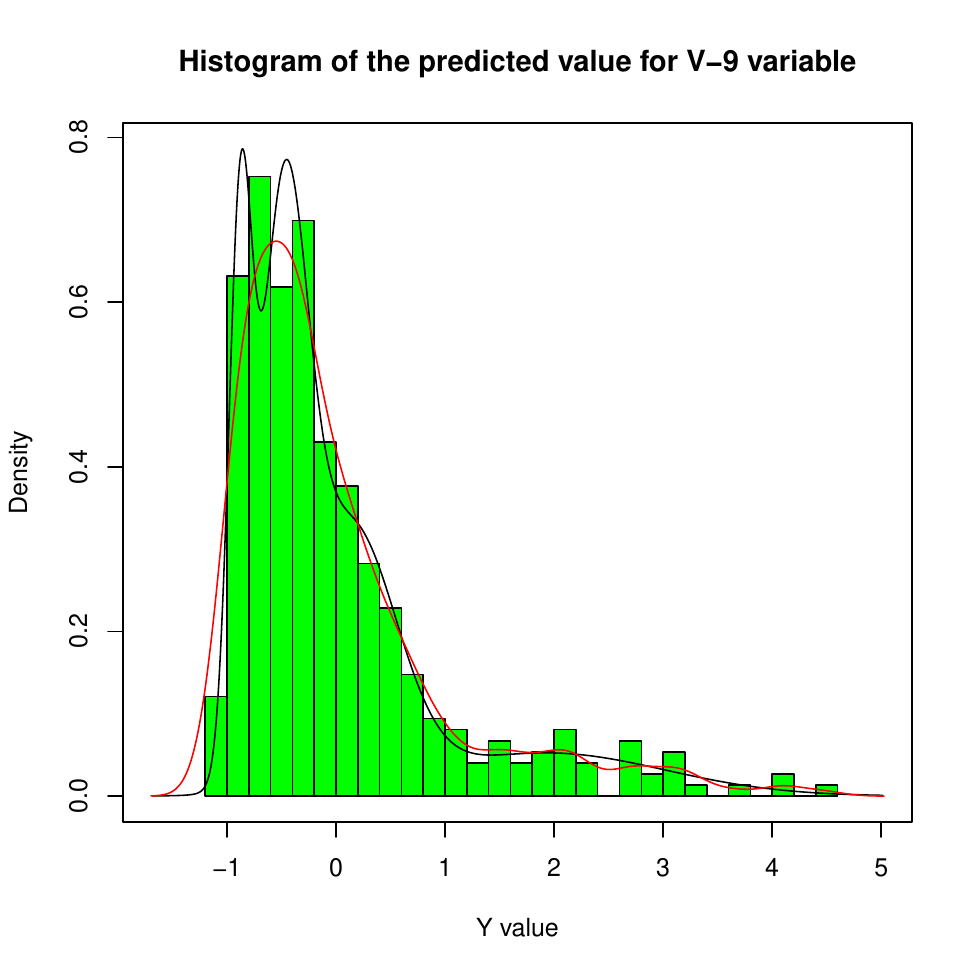}\\
 Histogram of $Y$ & Histogram of $\hat{Y}$ 
\end{tabular}
\caption{Histograms of the variable $Y$ (V-9) and its predicted value for residential building data. The red lines represent the estimated densities using non-parametric kernel density estimation, and the black lines show the densities estimated using a GMM.\label{RBDataHistogram}} 
\end{figure}
To evaluate the algorithm in a situation that has a moderate number of features, the Residential Building Data Set (UCI Machine Learning Repository) is used for further testing of the proximal Newton method in high-dimensional setting. 
This data set has $372$ observations and $108$ features, with the two response variables (V-9 and V-10), representing the sale prices and construction costs respectively. The V-9 variable (sale prices) is chosen as the response variable to be predicted.  Histogram of this variable is given in Figure \ref{RBDataHistogram}. As usual, all the features are standardized to have zero-mean and unit-variance. We provide the results of our algorithm with $K=3$ and $\lambda = 15$, $\gamma = 5$. 
The estimated parameters are publicly available on this link\footnote{\url{https://github.com/nv-thin/GLM-RMoE}}.

The correlation and the mean squared error between the true value V-9 with its prediction can be found in Table \ref{RBResult}.
These results show that the proximal Newton method performs well in this setting, in which it provides a sparse model and competitive criteria in prediction and clustering. The histogram of the predicted value $\hat{Y}$ in this case is close to the histogram of the true variable (see Figure \ref{RBDataHistogram}).
\begin{table}[!h]
    \centering
    \begin{tabular}{|c|c|c||c|c|c|c|c|}
\hline
& \multicolumn{2}{c||}{Predictive criteria}&\multicolumn{5}{c|}{Number of zero coefficients}\\
\cline{2-8}
Method & $R^2$ & MSE &  Exp.1 & Exp.2 & Exp.3 & Gate.1 & Gate.2\\
\hline
 GMoE-Lasso & 0.9994 & $0.00062_{(.0019)}$ & 71 & 38 & 75 & 97 & 101\\
\hline
\end{tabular}
\caption{Results for clustering the residential building data set.}
\label{RBResult}
\end{table}
\subsubsection{MoE model with Poisson outputs}
A data set is used here to illustrate the proposed regularized MoE of Poisson regression experts. We use the Cleveland Clinic Foundation heart disease data set, which is available from the UC Irvine Machine Learning Repository. The data set contains 13 features and 297 observations. The response variable is the heart disease diagnosis variable which takes integer values from 0 to 4 and is therefore treated as a count response in the Poisson regression experts. Among the 297 observations, 160 have a zero response value. Generally, an appropriate approach for such data with a large proportion of zero responses is to consider a zero-inflated Poisson regression model (ZIP model).
However, the regularized MoE of the Poisson regression is tested and observed on its behavior with this type of data. Taking $K=2$ and focusing on the regularized MoE for Poisson regression, the model's estimated parameters are provided in Table \ref{HeartDisease}. There are two components, the first one has 108 objects and the second one has 189 objects. The second class contains 156 over 160 observations that have zero response value. In this case, it looks like the data is splitted into two parts, with one part contains mainly zero response value similar with the approach of ZIP. In terms of prediction, $65\%$ of observations have the same values between their predictions and their response values. It is worth considering the regularized MoE for ZIP model as an extended approach for this type of data.
\begin{table}[!h]
\footnotesize
\begin{center}
\begin{tabular}{|c|c|c||c|}
\hline 
Feature & Exp.1 & Exp.2 & Gate\\
\hline
$x_{0}$	&0.51211	&-1.38996	&-0.71073\\
$x_{1}$	&-	&-	&-\\
$x_{2}$	&-	&-	&0.54763\\
$x_{3}$	&0.06753	&-	&0.54110\\
$x_{4}$	&0.00959	&0.09146	&-\\
$x_{5}$	&-	&-	&-\\
$x_{6}$	&-	&-	&-\\
$x_{7}$	&0.07229	&-	&0.10834\\
$x_{8}$	&-	&-	&-0.62335\\
$x_{9}$	&-	&0.50573	&-\\
$x_{10}$	&0.05960	&0.33149	&0.03440\\
$x_{11}$	&0.11976	&0.01285	&-\\
$x_{12}$	&0.05649	&-	&1.54824\\
$x_{13}$	&0.04244	&0.46287	&0.64450\\
\hline
\end{tabular}
\caption{Fitted regularized MoE model (PMoE-Lasso) to the heart disease data.\label{HeartDisease}}
\end{center}
\end{table}
\subsubsection{MoE model with Multinomial outputs}
For the multinomial case, we consider the two data sets that were used by \cite{Per14} in their work and compare the results between our approach with their method. We investigate the Ionosphere data and Musk-1 data which are described on the website UC Irvine Machine Learning Repository. The Ionosphere data contains $351$ observations and $33$ features. The Musk-1 data has $486$ observations and $168$ features. The variables with zero variance are removed. Hence, the Musk-1 data set remains with $167$ features. Both data sets have two classes. All features are standardized to have mean zero and unit variance. $K = 2$ is taken as in \cite{Per14}.\vk
The parameter estimates of the MoE models obtained by Lasso are given in Table \ref{Ionosphere} for Ionosphere data set. 
For Musk-1 data set, these parameters are publicly available on this link\footnote{\url{https://github.com/nv-thin/GLM-RMoE}}. The classification accuracy and percentage of features reduction results between the proposed method and Peralta's work are found in Table \ref{multinomial-results}. These results suggest that the proposed algorithm with Lasso provides better results than the remaining method in terms of data classification and features reduction. For Ionosphere dataset, Peralta used on average $78.1\%$ of all dimensions while our approach requires only $26.3\%$. For the Musk-1 dataset, the proposed Lasso method also improves the dimensionality-reduction rate by about 10 percentage points. Consider the classification rate, on both data sets the proposed method improves this rate by about 12 percentage points compared with Peralta's. One of the reasons for this improvement is that the approach of Peralta does not guarantee the increase of the penalized log-likelihood values after each loop of their EM algorithm.  
\begin{table}[htbp]
\small
\begin{center}
\begin{tabular}{|c|c|c|c||c|c|c|c|}
\hline 
Feature & Exp.1 & Exp.2 & Gate & Feature & Exp.1 & Exp.2 & Gate\\
\hline
$x_{0}$	&-1.64671	&-1.25999	&0.34349 &  $x_{17}$	&-	&-0.29576	&-\\
$x_{1}$	&-1.04171	&-0.79945	&- & $x_{18}$	&-	&-	&-\\
$x_{2}$	&-0.94925	&-0.64691	&- & $x_{19}$	&-	&-	&-\\
$x_{3}$	&-	&-	&- & $x_{20}$	&-	&-	&-\\
$x_{4}$	&-	&-1.81555	&0.94631 & $x_{21}$	&0.41903	&-	&-\\
$x_{5}$	&-0.05046	&-0.20732	&- & $x_{22}$	&-	&-	&-\\
$x_{6}$	&-0.45212	&-0.27119	&- & $x_{23}$	&-1.36138	&1.48880	&-1.83610\\
$x_{7}$	&-0.85935	&-0.18387	&- & $x_{24}$	&-0.41763	&-	&-\\
$x_{8}$	&-0.04429	&-	&- & $x_{25}$	&-	&-	&-\\
$x_{9}$	&-0.75204	&-	&-0.28020 & $x_{26}$	&-	&0.20319	&-\\
$x_{10}$	&-	&-	&- & $x_{27}$	&-	&-	&-\\
$x_{11}$	&-	&-	&- & $x_{28}$	&-	&-	&-\\
$x_{12}$	&-	&-	&- & $x_{29}$	&-	&-0.02892	&-\\
$x_{13}$	&-	&-	&- & $x_{30}$	&-	&-	&-\\
$x_{14}$	&-	&-	&- & $x_{31}$	&-	&-	&-\\
$x_{15}$	&-	&-0.15926	&- & $x_{32}$	&-	&-	&-\\
$x_{16}$	&-	&-	&- & $x_{33}$	&0.99009	&-0.21365	&-\\
\hline
\end{tabular}
\caption{Fitted regularized MoE model (MMoE-Lasso) to Ionosphere data.\label{Ionosphere}}
\end{center}
\end{table}
\begin{table}
\begin{center}
\begin{tabular}{|c|c|c||c|c|}
\hline 
Dataset name & \multicolumn{2}{c||}{Classification accuracy} & \multicolumn{2}{c|}{Dimensionality reduction}\\
\cline{2-5}
 & MMoE-Lasso (Peralta) & MMoE-Lasso & MMoE-Lasso (Peralta) & MMoE-Lasso\\
\hline
Ionosphere & $84.1\%$ & $96.6\%$ & $21.9\%$ & $73.7\%$\\
Musk-1 & $80.0\%$ & $93.3\%$ & $79.6\%$ & $90.0\%$\\
\hline
\end{tabular}
\caption{Classification accuracy and percentage of features reduction results compared between Peralta's method \citep{Per14} and our proposed method (MMoE-Lasso).\label{multinomial-results}}
\end{center}
\end{table}

\section{Conclusion and future work}\label{Sec:Con}

In this work, we proposed a regularized maximum-likelihood approach for the MoE
model that encourages sparsity, and developed EM-based algorithms that monotonically increase the regularized objective and converge to at least a local maximizer, without requiring the penalty approximations used in standard
regularized MoE procedures.
The proposed algorithms rely on proximal Newton-type methods and univariate parameter updates via coordinate ascent, which avoid matrix inversion and yield exactly sparse solutions.
On both simulated and real data sets, the results in terms of parameter estimation, recovery of the true sparsity support, and clustering accuracy confirm the effectiveness of the proposal: the induced sparsity does not introduce
substantial bias in parameter estimation, nor does it degrade the recovery of the underlying clusters of the heterogeneous data.
Within the coordinate-ascent M-step, the proximal Newton-type approximation admits closed-form coordinate-wise updates, making the method computationally efficient and applicable to moderate- and higher-dimensional data sets.
Future work may investigate more extensive model-selection experiments, the extension to hierarchical MoE of generalized linear models, and theoretical guarantees (e.g. consistency and support recovery) in genuinely high-dimensional regimes.


\appendix
\section{\label{appendixA}Proximal Newton-type methods}
Assume that we want to solve an optimization problem given by
\begin{equation*}
\min\limits_{x\in\mathbb{R}^n}f(x) = g(x) + h(x),
\end{equation*}
 with a composite function $f(x)$ where $g$ is a convex, continuously differentiable loss function, and $h$ is a convex but non differentiable penalty function. Such problems include the Lasso, elastic net, etc. Proximal Newton-type methods approximate only the smooth part $g$ with a local quadratic function of the form: 
\begin{equation}\label{prox-func1}
\hat{f}_k(x) = g(x_k) + \triangledown g(x_k)^\top(x-x_k) + \frac{1}{2}(x-x_k)^\top H_k(x-x_k)+ h(x),
\end{equation}
where $\triangledown g(x_k)$ is the gradient vector of $g$ at $x_k$ and $H_k$ is an approximation to the Hessian matrix $\triangledown^2g(x_k)$. If we choose $H_k=\triangledown^2g(x_k)$, we obtain the {\it proximal Newton method}. In this method, one uses an iterative algorithm with initial value $x_0$ and in which at step $k$ minimizes the proximal function $\hat{f}_k(x)$ instead of $f$ and then searches for the next value $x_{k+1}$ based on the solution of (\ref{prox-func1}) that will improve the value of $f$, i.e., $f(x_{k+1}) < f(x_k)$ by using a back tracking line search (see \cite{Boyd04}) until the algorithm converges. \cite{Lee14} and \cite{S.Lee06} studied convergence properties of proximal Newton methods. This strategy has some advantages especially since $g(x)$  does not have a quadratic form and $h(x)$ is the $\ell_1$ penalty function. First, by approximating $g(x)$ with its local quadratic form, several good methods can be used to solve (\ref{prox-func1}) such as coordinate ascent, where updating one parameter in each step will avoid computing the inverse of a matrix. Second, one can obtain the closed-form update for each parameter at each iteration of the algorithm, hence, reduce the computational time of the algorithm. Finally, for searching $x^{(t+1)}$, one can use the efficient backtracking line search strategy which is easy to setup.
A generic proximal Newton-type method can be listed as in Algorithm \ref{algo3} (see \cite{Lee14}).
\begin{algorithm}\label{algo:PN-type}
\caption{A generic proximal Newton-type procedure}
\label{algo3}
\begin{algorithmic}[1]
\STATE Starting point $x_0\in\text{dom}f$.
\REPEAT
\STATE Choose $H_k$, a positive definite approximation to the Hessian.
\STATE Solve the subproblem for a search direction:
$$\triangle x_k \leftarrow \arg\min_d  \triangledown g(x_k)^\top d + \frac{1}{2}d^\top H_kd+ h(x_k+d).$$
\STATE Select $t_k$ with a backtracking line search.
\STATE Update: $x_{k+1} \leftarrow x_k + t_k\triangle x_k$.
\UNTIL a stopping condition is satisfied.
\end{algorithmic}
\end{algorithm}
\section{Partial quadratic approximation for the gating network}\label{appendixB}
The $Q(\bw; \bstheta^{[q]})$ function in (\ref{QP}) is given as following
$$Q(\bw; \bstheta^{[q]}) = I(\bw) - \sum_{k=1}^{K-1}\gamma_k\|\bsw_k\|_1,$$
where the concave, continuously differentiable function $I(\bw)$ is 
$$I(\bw) = \sum_{i=1}^n\sum_{k=1}^{K-1}\tau_{ik}^{[q]}(w_{k0}+\bsx_i^\top  \bsw_k)- \sum_{i=1}^n\log\Bigl[1+\sum_{k=1}^{K-1}e^{w_{k0}+\bsx_i^\top  \bsw_k}\Bigl]$$
By taking the first and second derivatives of $I(\bw)$ w.r.t $(w_{k0},\bsw_k)$
\begin{align*}
\frac{\partial I(\bw)}{\partial w_{kj}} &= \sum\limits_{i=1}^n(\tau_{ik}^{[q]}-\pi_k(\bsx_i;\bw))x_{ij},\\
\frac{\partial^2 I(\bw)}{\partial w_{kj}\partial w_{kh}} &= -\sum\limits_{i=1}^n x_{ij}x_{ih}\pi_k(\bsx_i;\bw)(1-\pi_k(\bsx_i;\bw)),
\end{align*}
for $j, h \in \{0,1,\hdots, p\}$ with $x_{i0} = 1$, then the partial quadratic approximation to $I(\bw)$ w.r.t $(w_{k0}, \bsw_k)$ at $(\tilde{w}_{k0}, \tilde{\bsw}_k)$ is given by
\begin{equation*}
l_{I_k}(w_{k0},\bsw_k) = -\frac{1}{2}\sum\limits_{i=1}^nd_{ik}(c_{ik} - w_{k0} - \bsx_i^\top \bsw_k)^2 + C(\tilde{\bw}),
\end{equation*}
and
\begin{align*}
c_{ik} & = \tilde{w}_{k0} + \bsx_i^\top \tilde{\bsw}_k + \frac{\tau_{ik}^{[q]} - \pi_k(\tilde{\bw};\bsx_i)}{\pi_k(\tilde{\bw};\bsx_i)(1-\pi_k(\tilde{\bw};\bsx_i))},\\
d_{ik} & = \pi_k(\tilde{\bw};\bsx_i)(1-\pi_k(\tilde{\bw};\bsx_i)),
\end{align*}
$ C(\tilde{\bw})$ is a function of $\tilde{\bw}$. 

\section{The Quadratic Approximation for the Expert Network}
\label{app:quadratic_approx}

We derive the quadratic approximation \eqref{eq:quadratic_approx_expert} from the local quadratic model \eqref{eq:taylor_expansion} for the function $P_k(\boldsymbol{\theta}_k;\boldsymbol{\theta}^{[q]})$ defined in \eqref{eq:Q_expert_decomposed}.

Recall the partial log-likelihood for the $k$-th expert:
\begin{equation*}
P_k(\boldsymbol{\theta}_k;\boldsymbol{\theta}^{[q]}) = \sum_{i=1}^n \tau^{[q]}_{ik} \left[ \frac{y_i\eta_k(\boldsymbol{x}_i) - b(\eta_k(\boldsymbol{x}_i))}{a(\phi_k)} \right],
\end{equation*}
where $\eta_k(\boldsymbol{x}_i) = \beta_{k0} + \boldsymbol{x}_i^\top \boldsymbol{\beta}_k$.

In what follows, we consider $\tilde{\boldsymbol{\theta}}_k = (\tilde{\beta}_{k0}, \tilde{\boldsymbol{\beta}}_k^\top)^\top$ 
as the current estimate of the regression coefficients, 
treating the dispersion parameter $\phi_k$ as fixed for the optimization step, and $\tilde{\eta}_k(\boldsymbol{x}_i) = \tilde{\beta}_{k0} + \boldsymbol{x}_i^\top \tilde{\boldsymbol{\beta}}_k$.

\subsection*{Gradient and Hessian of $P_k$}

For a canonical link, we have:
\[
\frac{\partial}{\partial \eta_k} \bigl[ y_i\eta_k - b(\eta_k) \bigr] = y_i - b'(\eta_k),
\qquad
\frac{\partial^2}{\partial \eta_k^2} \bigl[ y_i\eta_k - b(\eta_k) \bigr] = - b''(\eta_k).
\]

Let $x_{i0}=1$ and $x_{ij}$ be the $j$-th predictor for observation $i$. Then:
\begin{align}
\frac{\partial P_k}{\partial \beta_{kj}} 
&= \sum_{i=1}^n \frac{\tau^{[q]}_{ik}}{a(\phi_k)} \bigl[ y_i - b'(\eta_k(\boldsymbol{x}_i)) \bigr] x_{ij}, \label{eq:gradient_Pk} \\
\frac{\partial^2 P_k}{\partial \beta_{kj} \partial \beta_{k\ell}} 
&= - \sum_{i=1}^n \frac{\tau^{[q]}_{ik}}{a(\phi_k)} b''(\eta_k(\boldsymbol{x}_i)) \, x_{ij} x_{i\ell}. \label{eq:hessian_Pk}
\end{align}

\subsection*{Second-order Taylor Expansion}

The quadratic approximation around $\tilde{\boldsymbol{\theta}}_k$ is:
\begin{equation}
\tilde{P}_k(\boldsymbol{\theta}_k) \approx P_k(\tilde{\boldsymbol{\theta}}_k) + \nabla P_k(\tilde{\boldsymbol{\theta}}_k)^\top (\boldsymbol{\theta}_k - \tilde{\boldsymbol{\theta}}_k) 
+ \frac12 (\boldsymbol{\theta}_k - \tilde{\boldsymbol{\theta}}_k)^\top \nabla^2 P_k(\tilde{\boldsymbol{\theta}}_k) (\boldsymbol{\theta}_k - \tilde{\boldsymbol{\theta}}_k).
\label{eq:taylor_expansion_app}
\end{equation}

Substituting \eqref{eq:gradient_Pk}--\eqref{eq:hessian_Pk} evaluated at $\tilde{\eta}_k(\boldsymbol{x}_i)$:
\begin{align*}
\nabla P_k(\tilde{\boldsymbol{\theta}}_k)^\top (\boldsymbol{\theta}_k - \tilde{\boldsymbol{\theta}}_k)
&= \sum_{i=1}^n \frac{\tau^{[q]}_{ik}}{a(\phi_k)} \bigl[ y_i - b'(\tilde{\eta}_k(\boldsymbol{x}_i)) \bigr] 
\bigl[ (\beta_{k0} - \tilde{\beta}_{k0}) + \boldsymbol{x}_i^\top (\boldsymbol{\beta}_k - \tilde{\boldsymbol{\beta}}_k) \bigr], \\
(\boldsymbol{\theta}_k - \tilde{\boldsymbol{\theta}}_k)^\top \nabla^2 P_k(\tilde{\boldsymbol{\theta}}_k) (\boldsymbol{\theta}_k - \tilde{\boldsymbol{\theta}}_k)
&= - \sum_{i=1}^n \frac{\tau^{[q]}_{ik}}{a(\phi_k)} b''(\tilde{\eta}_k(\boldsymbol{x}_i)) 
\bigl[ (\beta_{k0} - \tilde{\beta}_{k0}) + \boldsymbol{x}_i^\top (\boldsymbol{\beta}_k - \tilde{\boldsymbol{\beta}}_k) \bigr]^2.
\end{align*}

\subsection*{Reformulation as Weighted Least Squares}

Define the \emph{working response} and \emph{weights}:
\begin{equation*}
y^*_{ik} = \tilde{\eta}_k(\boldsymbol{x}_i) + \frac{y_i - b'(\tilde{\eta}_k(\boldsymbol{x}_i))}{b''(\tilde{\eta}_k(\boldsymbol{x}_i))},
\qquad
\omega_{ik} = \frac{\tau^{[q]}_{ik}}{a(\phi_k)} b''(\tilde{\eta}_k(\boldsymbol{x}_i)).
\end{equation*}

Then, after completing the square and collecting terms that depend only on $\tilde{\boldsymbol{\theta}}_k$ into a constant $C(\tilde{\boldsymbol{\theta}}_k)$, the expansion \eqref{eq:taylor_expansion_app} simplifies to:
\begin{equation}
\tilde{P}_k(\boldsymbol{\theta}_k) = -\frac{1}{2} \sum_{i=1}^n \omega_{ik} \bigl( y^*_{ik} - \beta_{k0} - \boldsymbol{x}_i^\top \boldsymbol{\beta}_k \bigr)^2 + C(\tilde{\boldsymbol{\theta}}_k).
\label{eq:weighted_least_squares}
\end{equation}
This matches equation \eqref{eq:quadratic_approx_expert} in the main text.

\subsection*{Verification}

Expanding the square:
\[
\bigl( y^*_{ik} - \beta_{k0} - \boldsymbol{x}_i^\top \boldsymbol{\beta}_k \bigr)^2
= \bigl[ (y^*_{ik} - \tilde{\eta}_k(\boldsymbol{x}_i)) - \bigl( (\beta_{k0} - \tilde{\beta}_{k0}) + \boldsymbol{x}_i^\top (\boldsymbol{\beta}_k - \tilde{\boldsymbol{\beta}}_k) \bigr) \bigr]^2.
\]

Noting that $y^*_{ik} - \tilde{\eta}_k(\boldsymbol{x}_i) = \dfrac{y_i - b'(\tilde{\eta}_k)}{b''(\tilde{\eta}_k)}$, and using the definitions of $\omega_{ik}$ and $y^*_{ik}$, one can verify that \eqref{eq:weighted_least_squares} is algebraically equivalent to \eqref{eq:taylor_expansion}.

Thus, the quadratic approximation of $P_k$ reduces to a penalized weighted least-squares problem, which can be efficiently solved via coordinate ascent with soft-thresholding for $\ell_1$-regularized estimation.

\section{Specific Forms of Working Responses and Weights}
\label{app:expert_forms}

This appendix provides the specific forms of the working response $y_{ik}^*$ and weights $\omega_{ik}$ for the three expert models considered in this work.

\subsection*{Gaussian Experts}

For Gaussian experts with identity link and fixed variance $\sigma_k^2$, we have:
\begin{equation*}
 a(\phi_k) = \sigma_k^2, \ b(\eta) = \eta^2 / 2,\ b'(\eta) = \eta,\ b''(\eta) = 1   
\end{equation*}

This leads to the simple values:
\[
y_{ik}^* = y_i, \quad \omega_{ik} = \frac{\tau_{ik}^{[q]}}{\sigma_k^2}.
\]
The update reduces to a weighted Lasso linear regression. After updating $(\beta_{k0}, \bm{\beta}_k)$, the variance is updated as:
\[
\sigma_k^{2[q+1]} = \frac{\sum_{i=1}^n \tau_{ik}^{[q]}(y_i - \beta_{k0}^{[q+1]} - \bm{x}_i^\top  \bm{\beta}_k^{[q+1]})^2}{\sum_{i=1}^n \tau_{ik}^{[q]}}.
\]

\subsection*{Poisson Experts}

For Poisson experts with log link, we have:
\begin{equation*}
 a(\phi_k) = 1, \ 
 b(\eta) = e^{\eta}, \ 
 b'(\eta) = e^{\eta}, \ 
 b''(\eta) = e^{\eta}   
\end{equation*}

The working response and weights are:
\[
y_{ik}^* = \tilde{\eta}_k(\bm{x}_i) + \frac{y_i - e^{\tilde{\eta}_k(\bm{x}_i)}}{e^{\tilde{\eta}_k(\bm{x}_i)}}, \quad \omega_{ik} = \tau_{ik}^{[q]} e^{\tilde{\eta}_k(\bm{x}_i)}.
\]

\subsection*{Multinomial Experts}

For a Multinomial expert with $R$ classes and a softmax gating network, the model for component probabilities is:
\[
\alpha_{kr}(\bm{x}_i; \bm{\beta}_k) = \mathbb{P}(y_i = r | \bm{x}_i; z_i = k) = \frac{\exp(\beta_{kr0} + \bm{x}_i^\top  \bm{\beta}_{kr})}{1 + \sum_{l=1}^{R-1} \exp(\beta_{kl0} + \bm{x}_i^\top  \bm{\beta}_{kl})}, \quad \text{for } r=1,\ldots,R-1.
\]
Let $u_{ir} = \mathbb{I}(y_i = r)$. The update for the parameters of a single class $r$ $(\beta_{kr0}, \bm{\beta}_{kr})$ follows the same logic as the gating network update in Section \ref{sec:GatingNetworkUpdating}. The working response and weights for class $r$ are:
\[
y_{ikr}^* = \tilde{\eta}_{kr}(\bm{x}_i) + \frac{u_{ir} - \alpha_{kr}(\bm{x}_i; \tilde{\bm{\beta}}_k)}{\alpha_{kr}(\bm{x}_i; \tilde{\bm{\beta}}_k)(1 - \alpha_{kr}(\bm{x}_i; \tilde{\bm{\beta}}_k))}, \quad \omega_{ikr} = \tau_{ik}^{[q]} \alpha_{kr}(\bm{x}_i; \tilde{\bm{\beta}}_k)(1 - \alpha_{kr}(\bm{x}_i; \tilde{\bm{\beta}}_k)).
\]
The parameters for each class are updated sequentially by solving a penalized weighted least squares problem of the form \eqref{eq:expert_lasso_problem}.
\renewcommand{\baselinestretch}{1.3}
\normalsize
\setlength{\bibsep}{0.1pt}

\bibliographystyle{apalike}
\bibliography{RMoE}

\end{document}